\documentclass[letterpaper,12pt]{article}
\usepackage{arxiv}
\usepackage[pdftex]{graphicx}
\graphicspath{{images/}}
\usepackage{pifont}
\usepackage{todonotes} 
\usepackage{url} 
\usepackage{color}
\usepackage{enumerate} 
\usepackage{array}
\usepackage{multirow}
\usepackage{float}
\usepackage{enumitem}
\usepackage{amsmath,amssymb}
\usepackage{verbatim}
\usepackage{footnote}
\usepackage{adjustbox}
\usepackage{multirow}
\usepackage{array}
\usepackage{makecell}
\usepackage{listings}
\usepackage{longtable}
\usepackage{threeparttable}
\usepackage{booktabs}
\usepackage{subcaption}

\newcolumntype{C}[1]{>{\centering\arraybackslash}m{#1}}
\graphicspath{{figures/}}

\definecolor{mygreen}{rgb}{0,0.6,0}
\definecolor{mygray}{rgb}{0.5,0.5,0.5}
\definecolor{mymauve}{rgb}{0.58,0,0.82}

\lstdefinestyle{default_listing}
{
    backgroundcolor=\color{white},           
    breakatwhitespace=false,                 
    breaklines=true,                         
    captionpos=b,                            
    commentstyle=\color{mygreen},            
    escapeinside={(*@}{@*)},                 
    extendedchars=true,                      
    frame=tb,                                
    float=!th,
    keepspaces=true,                         
    columns=fullflexible,
    keywordstyle=\color{blue},               
    numbers=left,                            
    numbersep=5pt,                           
    numberstyle=\scriptsize\color{mygray},  
    rulecolor=\color{black},                 
    showspaces=false,                        
    showstringspaces=false,                  
    showtabs=false,                          
    stepnumber=1,                            
    stringstyle=\color{mymauve},             
    tabsize=2,                               
    xleftmargin=1.5em,
    framexleftmargin=1em
}

\lstdefinestyle{java}
{
    style=default_listing,
    language=Java
}

\lstdefinelanguage{JavaScript}{
  keywords={break, case, catch, continue, debugger, default, delete, do, else, finally, for, function, if, in, instanceof, new, return, switch, this, throw, try, typeof, var, void, while, with},
  morecomment=[l]{//},
  morecomment=[s]{/*}{*/},
  morestring=[b]',
  morestring=[b]",
  sensitive=true
}

\lstdefinestyle{javascript}
{
    style=default_listing,
    language=JavaScript
}

\lstnewenvironment{envcode}[1]{
    
    \lstset{
        #1
    }
}{}

\newcolumntype{P}[1]{>{\centering\arraybackslash}p{#1}}
\newcolumntype{M}[1]{>{\centering\arraybackslash}m{#1}}

\newcommand\tool{\textsc{BrenntDroid}}

\begin{document}


\title{Dissecting Android Cryptocurrency Miners}

\author{
{\rm Stanislav Dashevskyi}\\
Forescout Technologies
\and
{\rm Yury Zhauniarovich}\\
Independent Researcher
\and
{\rm Olga Gadyatskaya}\\
LIACS\\
Leiden University
\and
{\rm Aleksandr Pilgun}\\
University of \\Luxembourg
\and
{\rm Hamza Ouhssain}\\
ARHS Developments
}

\maketitle


\subsection*{Abstract}

Cryptojacking applications pose a serious threat to mobile devices. Due to the extensive computations, they deplete the battery fast and can even damage the device. In this work we make a step towards combating this threat. We collected and manually verified a large dataset of Android mining apps. In this paper, we analyze the gathered miners and identify how they work, what are the most popular libraries and APIs used to facilitate their development, and what static features are typical for this class of applications. Further, we analyzed our dataset using VirusTotal. The majority of our samples is considered malicious by at least one VirusTotal scanner, but 16 apps are not detected by any engine; and at least 5 apks were not seen previously by the service.

Mining code could be obfuscated or fetched at runtime, and there are many confusing miner-related apps that actually do not mine. Thus, static features alone are not sufficient for miner detection. We have collected a feature set of dynamic metrics both for miners and unrelated benign apps, and built a machine learning-based tool for dynamic detection. Our \tool\ tool is able to detect miners with 95\% of accuracy on our dataset. 

This preprint is a technical report accompanying the paper ``Dissecting Android Cryptocurrency Miners'' published in ACM CODASPY 2020~\cite{dashevskyi2020dissecting}.

\section{Introduction}

The recent wave of cryptocurrencies contributed to the debut of a new malware class called \emph{cryptominers}, \emph{cryptojackers}, or simply \emph{miners}. After infecting a device, these malicious applications start solving computationally hard puzzles that support the cryptocurrency network, getting rewards for their work that are accumulated on the miner developer's account. The ease of monetization and the anonymity factors enabled the quick growth of the mining malware. In 2017, the skyrocketing price of cryptocurrencies caused by the enormous attention to these technologies has played a role in the cryptojacking proliferation~\cite{fireeyereport2018}. Not surprisingly, miners have quickly gained popularity and appeared among the top security threats in 2018~\cite{cyberthreatreport2018,sophos2018}. Security researchers have also paid attention, with many papers focusing on browser-based and binary mining~\cite{hong2018you,eskandari2018first,musch2018web,ruth2018digging,konoth2018minesweeper,saad2018end,papadopoulos2018truth,rauchberger2018other,pastrana2019first} appearing recently.

Due to the cryptocurrencies boom, end-user demand for mining applications has emerged. Prior to July 2018 everybody could simply find mining applications on Google Play and attempt to  generate a few cryptocoins on the mobile device. Yet, as the smartphone-based mining no longer generates interesting profits for the benign user~\cite{saad2018end,clay2018power}, the interest to these apps has significantly diminished. Google has removed mining apps from Google Play, but they are still available on alternative markets. The ``crash'' of the cryptocurrency market in the end of 2018 forced the operation of several mining services to shut down. For instance, the popular service CoinHive has announced discontinuation of its service in February 2019~\cite{CoinHive_Discountinuation}. Still, there are many alternatives, like CryptoLoot and JSEcoint, which are important cyber threats today\footnote{\scriptsize{\url{https://www.helpnetsecurity.com/2019/04/10/cryptomining-still-dominates/}}} and that may further proliferate during the next crypto boom. Due to the rising price of the Monero coin, in Summer 2019 the cryptomining malware was revitalized\footnote{\scriptsize{\url{https://www.zdnet.com/article/crypto-mining-malware-saw-new-life-over-the-summer-as-monero-value-tripled/}}}.

The Android ecosystem itself is huge, comprising not only mobile devices but also wearable technology, TVs and cars. It is therefore a lucrative target for adversaries due to the large number of potential victims. Even smart TV appliances can now be infected with mining Android apps\footnote{\scriptsize{\url{http://blog.netlab.360.com/adb-miner-more-information-en/}}}. Security industry reports mention that mining capabilities are being introduced to existing malware families or added into repackaged Android applications~\cite{cyberthreatreport2018,sophos2018,fireeyereport2018}. Attackers are constantly looking into different ways to deliver malware to Android-based devices. For instance, recently cryptomining malware was distributed in 21 different countries via open Android Debug Bridge (ADB) ports\footnote{\scriptsize{\url{https://blog.trendmicro.com/trendlabs-security-intelligence/cryptocurrency-mining-botnet-arrives-through-adb-and-spreads-through-ssh/}}}.

Indeed, mobile mining has certain advantages for the attackers. By running in the background and when the user is not present, the mining code packaged as an app can be \emph{more persistent} than website visits. The cost of creating and distributing a miner is \emph{negligible}, given the ease of application repackaging~\cite{salem2018repackman} on Android and the availability of mining libraries~\cite{konoth2018minesweeper}. But miners are particularly \emph{dangerous} for mobile devices. The extensive computations performed during the mining process drain the battery and increase the temperature of the device, potentially causing irreversible damage. For example, a malicious family called \texttt{Loapi} causes the mobile device's battery to overcook within 48 hours after the infection~\cite{kaspersky-loapi-2017}. Therefore, there is a need to study the Android miner phenomenon and to be able to detect such applications.

\paragraph{Contributions.}
To the best of our knowledge, our paper presents the first large study of Android miners. We make the following contributions.
\begin{itemize}[leftmargin=*]
	\item We have collected a large dataset of mining Android apps, which includes both Web-based and binary-based cryptocurrency miners. As our focus is on the Android mining phenomenon, our dataset contains  malicious mining applications, and also honest miners that declare their mining activity upfront and could be solicited by the users. We also include properly labelled samples of non-mining applications that can confuse the basic detection approaches (scam, wallet apps, etc.). Our dataset has been fully confirmed by manual analysis. We share our labelled dataset and the metadata with the community\footnote{\scriptsize{The dataset is available upon request at \url{https://standash.github.io/android-miners-dataset/}}}.

	\item We share insights on how (\emph{JavaScript}) and \emph{binary} Android miners work, how the mining code is injected, and what are the most popular libraries/APIs for mining. Particularly, we have identified \textbf{8} common mining libraries that are used in \textbf{671} miners.
	
	\item At least \textbf{5} miners from our dataset have previously never been uploaded to the popular VirusTotal service. We have also found \textbf{16} apps, including both malicious and honest miners, that are not detected by any VirusTotal scanner. Finally, we have ranked the antivirus engines at VirusTotal based on our dataset.    

	\item Using our verified dataset as the ground truth, we performed dynamic analysis of the miners and compared the results with randomly selected benign applications. We identified a set of dynamic metrics that are the most efficient for accurate classification results, achieving 95\% of accuracy and the AUC score of 0.988$\pm$0.009. Based on our findings, we propose the \tool\ tool that can be used to detect miners dynamically and to check if an app indeed mines cryptocurrencies. 
\end{itemize}

\section{Background}\label{sec:background}
In this section we provide the necessary background information about Android applications, and we detail how cryptomining can be executed by an Android app.
\subsection{Android Applications}

Android apps are distributed as archive files (\texttt{.apk})
that consist of the compiled Java code (\texttt{.dex}), the application manifest (\texttt{AndroidManifest.xml}) containing the requested Android permissions and subscriptions to system events, application resource files (e.g.,
Web-based resources such as \texttt{.js} and \texttt{.html} files), and native libraries (\texttt{.so} files).

The contents of \texttt{.apk} archives can be extracted with \emph{apktool}~\cite{apktool_webpage} that, among other
things, disassembles the \texttt{.dex} files and transforms the disassembled Java classes into \emph{smali}~\cite{smali}
-- a low-level human-readable code representation for Java code in Android platform.

Android has wide support for embedding Web-based content into its applications~\cite{luo2011attacks}. To achieve this,
Android apps support \texttt{WebView} -- a technology that packages the basic functionality of web browsers (e.g.,
webpage rendering and JavaScript) into a Java class that can be instantiated within an Android app and function
similarly to a basic web browser. In particular, this technology allows the developers of Android apps to embed Web
content into the application resources and interact with this content from the Java code and vice-versa.

\begin{envcode}{style=java, label={lst:webview-example}, caption={WebView example}}
WebView view = new WebView(this);
view.getSettings().setJavaScriptEnabled(true);
view.loadUrl("https://google.com");
view.loadUrl("file://assets/page.html");
\end{envcode}

The Android \texttt{WebView} is a subclass of the standard \texttt{View} class; its basic usage is shown in Listing
\ref{lst:webview-example}: line 1 instantiates an object of the \texttt{WebView} class, line 2 enables the execution of
JavaScript code (disabled by default), and lines 3 and 4 load Web-based content into the \texttt{WebView} element. Note
that not only the remote Web content can be loaded (line 3), but also the resource files shipped together with the app
(line 4).

\begin{envcode}{style=java, label={lst:native-lib-example}, caption={Native library load example}}
public class NativeLibrary {
    static {
        System.loadLibrary("native_lib");
    }
    public static native void doSmthng(int param);
}
\end{envcode}

\begin{envcode}{style=java, label={lst:executable-example}, caption={Running a shell command from an Android app}}
String command = "echo 'hello world'";
Runtime localRuntime = Runtime.getRuntime();
localRuntime.exec(command);
\end{envcode}

Listing~\ref{lst:native-lib-example} shows how a native library can be called via the \texttt{System.loadLibrary(...)}
interface: a Java wrapper class has to be created in the Android app, where the \texttt{.so} library has to be loaded
(line 3) and the corresponding native methods declared (line 5). Listing~\ref{lst:executable-example} illustrates how a
shell command can be executed from the Java code of an Android app.

\subsection{Android Cryptocurrency Miners}

There exist two different approaches for mining cryptocurrencies on mobile devices: (1) the mining code is embedded into a Web page that
can be executed via a Web browser (we refer to them as \emph{javascript} miners from now on); (2) the mining code is
packed into a binary that can be executed by a device (we will call them \emph{binary} miners). 

Both of the two mining approaches can be used to create either \emph{legitimate} or \emph{illicit} miners: the miners that
belong to the former category explicitly ask the user consent for mining, while the latter attempt to hide the fact that
they are mining. Also, the Web-based approach is typically used by the authors of malicious websites,
while the binary approach is favored by the authors of the traditional computer malware.  It is important to stress,
that both of these approaches can be used within Android apps.

Software that mines cryptocurrencies typically rely upon drive-by mining services such as
\emph{CoinHive}\footnote{https://krebsonsecurity.com/2018/03/who-and-what-is-coinhive/} that provide the necessary
infrastructure to mine cryptocurrencies such as Monero. This is particularly attractive for regular users, as Monero can
be mined using the CPU, instead of expensive GPU or other specialized hardware~\cite{konoth2018minesweeper}. There exist
other ``lightweight'' cryptocurrencies, such as Litecoin and Ethereum that can be mined using the commodity hardware.
However, according to various reports, Monero significantly dominates
them~\cite{cyberthreatreport2018,saad2018end,eskandari2018first}.  Also, to facilitate cryptocurrency mining with
comparatively weak hardware, mining service providers support creating \emph{mining pools}, when several devices combine
their computational power to perform mining collectively. Therefore, when multiple devices are combined into a single
mining pool, mining the lightweight cryptocurrencies becomes profitable.

\begin{envcode}{style=javascript, label={lst:js-orchestrator-example}, caption={JavaScript miner initalization example}}
<script src="https://coinhive.com/lib/coinhive.min.js"/>
<script>
    var miner = new CoinHive.Anonymous('SITE_KEY');
    miner.start();
</script>
\end{envcode}

Listing~\ref{lst:js-orchestrator-example} shows an example of an initialization script that, when embedded into a Web
page, starts mining the Monero cryptocurrency once that page is loaded. The ``\texttt{SITE\_KEY}'' needs to be
substituted by the actual hash of the public site key, which is connected to a certain cryptocurrency wallet tat willh
receive the mining reward (the \emph{mining credentials}). In this example, the \emph{anonymous} version of the CoinHive
API has been used: any device that loads the script will perform the mining, however the reward will be received only by
the owner of the site key. Moreover, a wallet can have multiple public site keys associated to it, and the ``identity''
of the wallet behind the miner cannot be inferred. Such a simple mining initalization script is particularly popular in
\emph{malicious} website mining, as it is takes no effort for embedding it, and provides anonymity~\cite{hong2018you}.

\begin{envcode}{style=default_listing, label={lst:bin-miner-example}, caption={Binary miner initalization example}}
./minerd --url stratum+tcp://eu.multipool.us:7777 --userpass USERNAME:PASSWORD
\end{envcode}

Similarly, the script on Listing~\ref{lst:js-orchestrator-example} can be invoked via the \emph{WebView} element that
supports loading Web pages and JavaScript code in Android apps. Therefore, Android apps can use the same mechanism for
mining crypto as regular websites.

Listing~\ref{lst:bin-miner-example} shows an example of invoking a binary miner from a shell (a
\emph{MinerD}\footnote{The source code is available at \url{https://github.com/pooler/cpuminer}} executable). Android
apps can also include such miner executables and call them in this fashion by spawning a separate application process.

\subsection{Our Terminology}
To summarize, in this paper, we distinguish the following categories of cryptocurrency miners: 
\begin{itemize}
	\item \emph{Illicit miners} are applications that try to perform \emph{stealth} mining, i.e. they do not warn users explicitly about the mining process, or \emph{selfish} mining, i.e., they mine cryptocurrencies explicitly, but redirect the profits to the attacker, not the user.
	\item \emph{Legitimate} or \emph{benign} miners are applications that both declare their mining activities and the user is the sole benefactor of the mining process.

	\item \emph{Miner-related} apps are applications that match many mining string patterns, but do not perform cryptocurrency mining (e.g., wallet apps).
    \item \emph{Scam} miners are applications that pretend to perform mining activity, but do not actually perform any mining.
    \item \emph{Javascript} miners are applications that include the web-based mining payload.

    \item \emph{Binary} miners are applications that include binary mining payload.
\end{itemize}


\section{Dataset Collection}%
\label{sec:dataset}

We started by collecting several samples of Android miners. These applications have been found based on relevant security industry blog posts and whitepapers, e.g., the SophosLabs report~\cite{sophos2018}, that had explicitly mentioned the hashes of
illicit Android miners. We have also collected several miners from different Android app stores (Google
Play\footnote{This research had started before Google decided to remove all mining apps from Google Play on $07/27/2018$.}, F-droid, etc.). This initial dataset has been used to create a set of strings and rules in the YARA notation\footnote{\url{http://virustotal.github.io/yara/}} indicating mining payload in the code and metadata. 

The initial set of miner-related strings and YARA rules contained only a few generic keywords, such as
``\texttt{Monero}'', ``\texttt{Litecoin}'', generic mining API calls such as \texttt{CoinHive.Anonymous()},
\texttt{CRLT.Anonymous()}, and \texttt{CoinHive.User()}, and domain names of the popular mining pools such as
``\texttt{us.litecoinpool.org}'' ``\texttt{mine.xmrpool.net}''. Yet, while such strings are already useful for finding
\emph{some} potential miners, they are insufficient. 

During already the first iterations of the dataset collection, we realized that a fully automatic search against many
diverse apps is inevitably prone to errors. This was unacceptable, as we aimed to build a reliable collection of apps
that could be used as the ground truth for detecting Android miners.
Therefore, we manually analyzed each app with at least a single match to the miner-related strings. We were looking for
characteristics of the mining activity and the intended interactions with user: (a) the mining code (e.g., the code that
initializes mining and the mining libraries); (b) how the mining can be triggered by a user (by interacting with an app
in a certain way or by simply launching its main activity); (c) supported cryptocurrencies; (d) the declared
functionality of the app and whether it tries to ``hide'' its intentions. We also aimed to find more string patterns
that can be used to extend our sample of miners. Once we had found a new pattern, we added it to our set of
miner-related strings, and re-ran the string search against the apps that had no previous matches and a new batch of
apps that we have been downloading. 

To summarize, we performed many iterations of the following steps: \texttt{(1)} find a large sample
of \emph{potential} Android miners using string search, and download them; \texttt{(2)} perform a manual analysis to
find the evidence that these apps are miners, and discard false-positives; \texttt{(3)} update the search strings
used at the step \texttt{(1)} with new patterns discovered at the step \texttt{(2)}. We repeated these steps multiple
times, increasing our dataset of Android miners and improving its quality. 

\subsection{Main Application Sources}

To find a large set of \emph{potential} miners, we used the popular platforms
\emph{VirusTotal}\footnote{\url{https://www.virustotal.com/}} and
\emph{Koodous}\footnote{\url{https://koodous.com/}}.  Services provided by these platforms are quite
different, but they both allow to search for Android apps that match specific criteria, and to download them. Below, we
briefly introduce these platforms and describe how we utilized them.

\emph{VirusTotal} checks user-submitted binaries (including Android apps) against several popular anti-virus
engines. Currently, \emph{VirusTotal} allows not only to perform scans in the black-box manner, but also to search over
the dynamic data of apps (e.g., the URLs that apps try to connect to), and to perform string pattern search over its application database. 

We first used the \emph{Private
API}\footnote{\url{https://www.virustotal.com/en/documentation/private-api}} of \emph{VirusTotal} to search
and download the apps from our original dataset by their hashes, and to collect their metadata. From this metadata we
extracted the information about the malware families that these apps potentially correspond to, and manually compiled a
list of the families related to cryptocurrency mining (as reported by the anti-virus engines used by \emph{VirusTotal}).
Next, we used the \emph{file search
functionality}\footnote{\url{https://www.virustotal.com/intelligence/help/file-search/}} of
\emph{VirusTotal}, and downloaded all Android apps that have been recently detected by at least one of the anti-virus
engines and belong to at least one malware family from our list. Additionally, we checked the dynamic information of
apps against a set of known miner-related strings (e.g., mining pools and domain names listed
in~\cite{konoth2018minesweeper, hong2018you}), and downloaded the matching apps as well.  

\emph{Koodous} is collaborative platform that allows to download Android apps, analyze them, and share the analysis
results. \emph{Koodous} performs static and dynamic analyses of apps using the state-of-art Android analysis tools like
Androguard\footnote{\url{https://github.com/androguard/androguard}} and the Cuckoo
sandbox\footnote{\url{https://cuckoosandbox.org/}}. Users can also write custom YARA
rules\footnote{\url{https://yara.readthedocs.io/en/v3.4.0/writingrules.html}} for finding and downloading
the matching apps. We used this functionality to create our own YARA rulesets for
searching and downloading potential Android miners from \emph{Koodous}, based on the set of miner-related strings we
identified during our manual analysis over the sample retrieved from \emph{VirusTotal} (see Section~\ref{sec:triage}).
We also searched for the YARA rules that had been already written by the community to detect miner apps, and
incorporated them into our ruleset as well.

\subsection{Manual Analysis}%
\label{sec:triage}

To confirm that an app is a cryptocurrency miner we performed its thorough manual analysis.  During the manual analysis
we treated every app as follows: we decompiled it with \emph{apktool}, matched the set of miner-related strings against
the decompiled files, and examined the app starting from the files where we found matches.  We paid special attention to
the resource files with extensions \texttt{.html}, \texttt{.js}, and \texttt{.xml}, native libraries (\texttt{.so}), and
executable files shipped with the app. Once we had located the \emph{mining initialization code} (e.g., a JavaScript
code fragment that inserts the mining credentials into a mining library and starts the mining, or a \emph{smali} code
fragment that calls a native mining library) and/or the mining code (e.g., a library that implements the mining
functionality), we looked for the entry points in the app that triggered the mining. For example, we found at least
\textbf{22} cases when the mining initialization code is placed directly into the \emph{MainActivity} class, or located
in a subclass of the \emph{Application} class -- in such cases, the mining starts immediately upon the app startup.
While performing this process, we have identified and collected other static indicators that suggest that the app under
question is a miner. We describe them in more detail in Section \ref{subsec:static-detection}.

Using the above static indicators (see Section \ref{subsec:static-detection} for more details), we have downloaded in
total 17159 Android apps using both \emph{VirusTotal} and \emph{Koodous}. After the manual analysis step, we obtained
the dataset of \textbf{728} Android miners (we describe the dataset in Section \ref{sec:sample_statistics}). During the collection phase, we might have missed miners, e.g., if they used advanced hiding techniques like dynamic code updates~\cite{StaDynA_Zhauniarovich2015}. Thus, dataset may be incomplete. However, to the best of our knowledge, currently this is the only publicly available dataset of mobile miners.


\section{Dataset Description}
\label{sec:sample_statistics}

\paragraph{Javascript vs binary miners.} Table~\ref{tab:miner-sample} lists the percentages of Android apps in our
dataset: the proportions of \emph{javascript} and \emph{binary} miners in the sample, and the numbers of illicit miners
among them. We also identified a small subset of \emph{miner-related} apps that are similar to miners, but do not
contain any mining code, and can be points of confusion for automatic miner detection approaches (we discuss them in
more detail further in this Section).  The distribution of apps in Table~\ref{tab:miner-sample} suggests that the most
popular way to create mining apps is with JavaScript, and that the majority of the miners in our sample are
\emph{illicit} (\textbf{614} illicit miners).

\begin{table}[!ht]
\centering
\scriptsize
\begin{tabular}{lr}
    \toprule
    \textbf{Category}                   & \textbf{\# samples (\%)}\\
    \midrule
    JavaScript                          & 594 (77.95\%)    \\
    \emph{JavaScript illicit}           & 563 (73.88\%)    \\
    Binary                              & 134 (17.59\%)    \\
    \emph{Binary illicit}               &  51 (6.69\%)    \\
    Miner-related                       &  34 (4.46\%)    \\
    \midrule
    \midrule
    Total                               & 762 \\
    \bottomrule
     & \\ 
\end{tabular}
\caption{The distribution of Android apps in our sample}
\label{tab:miner-sample}
\end{table}

\begin{table}[!t]
\footnotesize
		\centering
		\begin{tabular}{lcr}
			\toprule
			\textbf{Android permission}   &  \textbf{Protection level}                & \textbf{\#Miners (\%)} \\
			\midrule
			INTERNET                      &  N (D if API$<$23)  & 728 (100.00\%)   \\
			ACCESS\_NETWORK\_STATE        &  N  & 374 (51.37\%)    \\
			WAKE\_LOCK                    &  N (D if API$<$17)  & 351 (48.21\%)    \\
			WRITE\_EXTERNAL\_STORAGE      &  D  & 300 (41.21\%)    \\
			RECEIVE\_BOOT\_COMPLETED      &  N  & 258 (35.44\%)    \\
			READ\_EXTERNAL\_STORAGE       &  D (N if API$<$23)  & 203 (27.88\%)    \\
			c2dm.permission.RECEIVE       &  N  & 149 (20.47\%)    \\
			ACCESS\_WIFI\_STATE           &  N  & 138 (18.96\%)    \\
			VIBRATE                       &  N  & 135 (18.54\%)    \\
			READ\_PHONE\_STATE            &  D  & 128 (17.58\%)    \\
			\bottomrule
		\end{tabular}
		\caption{Top 10 Android permissions used by miners (Protection levels: D - dangerous, N - normal)}
		\label{tab:top-perms}
\end{table}

\begin{table}[!t]
		\centering
		\footnotesize
		\begin{tabular}{lr}
			\toprule
			\textbf{Android system event}                                & \textbf{\#Miners (\%)}\\
			\midrule
			android.intent.action.BOOT\_COMPLETED                           & 536 (73.63\%)    \\
            android.intent.action.QUICKBOOT\_POWERON                        & 169 (22.18\%)    \\
			android.intent.action.MY\_PACKAGE\_REPLACED                     & 161 (22.11\%)    \\
			com.android.vending.INSTALL\_REFERRER                           & 159 (21.84\%)    \\
			com.google.android.c2dm.intent.RECEIVE                          & 146 (20.05\%)    \\
			com.htc.intent.action.QUICKBOOT\_POWERON                        & 122 (16.76\%)    \\
			android.net.conn.CONNECTIVITY\_CHANGE                           & 95  (13.05\%)    \\
			android.intent.action.ACTION\_POWER\_DISCONNECTED               & 84  (11.26\%)    \\
			android.intent.action.ACTION\_POWER\_CONNECTED                  & 84  (11.26\%)    \\
			android.intent.action.BATTERY\_LOW                              & 68  ( 8.52\%)    \\
			\bottomrule
		\end{tabular}
		\caption{Top 10 system event subscriptions by miners}
		\label{tab:top-actions}
\end{table}

\paragraph{Top Android permissions and system events used.} 
Permissions and system events subscriptions are widely used as features to detect Android
malware~\cite{arp2014drebin,tam2017evolution}, and it is interesting to see whether the miner population uses the same
permissions and listens to the same system events as generic malware.  

Wang et al.~\cite{wang2014exploring}, Jiang and Zhou~\cite{jiang2012dissecting}, and Feldman et
al.~\cite{feldman2014manilyzer} have previously compared Android malware and benign apps in terms of requested
permissions.  Table \ref{tab:top-perms} lists the top 10 requested Android permissions across our sample of miners. It
is evident that the only permission needed for Android miners to properly function is the
\texttt{android.permission.INTERNET}, which is the only permission requested by \textbf{286} miners. This permission is
not considered dangerous anymore, and is granted by the Android system without user
consent~\cite{zhauniarovich2016small}. Comparing the statistics in the works of Wang et al.~\cite{wang2014exploring} and
Jiang and Zhou\cite{jiang2012dissecting} with Table~\ref{tab:top-perms}, we can conclude that miners generally do not
request permissions that are very prevalent in malicious samples only, e.g., \texttt{READ\_SMS}. 

Table~\ref{tab:top-actions} lists the top 10 most occurring system event subscriptions, else called filtered intents.
Most of the miners have subscribed to \texttt{android.intent.action.BOOT\_COMPLETED}, which means that they will attempt
to resume their work as soon as the device has been booted. This system event is highly indicative of malicious
apps~\cite{zhu2016featuresmith}.  We see also that a significant number of miners tries to monitor the battery
consumption and the network connection status. 

Only \textbf{39} miners from our sample can be characterized as generic malware with respect to their behavior, while
\textbf{30} of these miners are actively forcing users to allow them administrative privileges, and monitor whether they
receive the role of the device administrator, and whether users try to revoke this role.

\begin{table*}[!ht]
\centering
\scriptsize
\begin{tabular}{llr}
    \toprule
    \textbf{Library}        & \textbf{URL}                                               & \textbf{\#Apps} \\
    \midrule
    CoinHive Android SDK    & \url{https://github.com/theapache64/coin_hive_android_sdk} & 437           \\
    CoinHive API            & \url{https://coinhive.com/lib/coinhive.min.js}             & 139           \\
    CPUMiner                & \url{https://github.com/pooler/cpuminer}                   & 42            \\
                            & \url{https://github.com/mdelling/cpuminer-android}         &               \\
    MinerD                  & \url{https://github.com/MiniblockchainProject/Minerd}      & 26            \\
    CGMiner                 & \url{https://github.com/reorder/cgminer_keccak}            & 17            \\
                            & \url{https://github.com/Max-Coin/cgminer}                  &               \\
    XMRig                   & \url{https://github.com/xmrig/xmrig}                       & 6             \\
    Authedmine API          & \url{https://authedmine.com/media/miner.html}              & 3             \\
    C0nw0nk                 & \url{https://github.com/C0nw0nk/CoinHive}                  & 1             \\
    \midrule
    UNKNOWN                 &                                                            & 57            \\
    \bottomrule
\end{tabular}
\caption{Third-party mining libraries used by the miners from our sample}
\label{tab:libraries}
\end{table*}

\paragraph{Mining libraries.}
\textbf{8} third-party mining libraries have been identified in our sample. Table \ref{tab:libraries}
lists these libraries and the number of apps from our sample that rely on them. In total, these libraries are used by
\textbf{671} miners. We could not identify the origin of the mining code for the remaining \textbf{57} miners. This was
either due to the fact that they might be using a custom mining library that we could not identify (mostly the case for
\emph{legitimate} miners), or the library has been heavily changed and obfuscated so that we could not match it to any of
the original libraries (mostly the case for \emph{illicit} miners). 

Notably, for \emph{javascript} miners, the plain JavaScript \emph{CoinHive API}\footnote{An example of its usage is
shown in Listing~\ref{lst:js-orchestrator-example}} library is \emph{not} the most used one: \textbf{437} miners
integrate the \emph{CoinHive Android SDK} library, which is a wrapper that provides convenient JavaScript-to-Java
bindings for the CoinHive API in Android apps.  At the same time, we found that the  \emph{Authedmine}\footnote{This
API has been released by CoinHive as well. Unlike the original mining API, it requires explicit user consent for
mining.} API has been used only in \textbf{3} cases.  We identified the usage of several desktop cryptomining software
projects in \emph{binary} miners: \emph{CPUMiner}, \emph{CGMiner}, \emph{XMRig}, and \emph{MinerD}. These projects have
been specifically compiled for Android as libraries/executables by the authors of the miners. We found several versions
of \emph{CPUMiner} and \emph{CGMiner} available on GitHub\footnote{\url{https://github.com}} used by the miners.  

We observed that in many cases the third-party libraries have been used ``as is'', however in some cases the original
library is changed by the authors of miners.  For instance, the original \emph{CoinHive Android SDK} library has had
large modifications in at least \textbf{64} \emph{illicit} miners from our sample. In all these cases the changes were
non-significant to the core functionality of the library (perhaps, made only for evading detection): e.g., package name
has been changed, several classes not related to mining have been removed, variable names have been changed, etc. For
example, in \textbf{8} of these cases, the ``\texttt{engine.html}'' file, which is the core of the library, has been
totally unchanged (we checked the hashes of the files against the original file provided by the library). In \textbf{56}
cases the ``\texttt{engine.html}'' file has been renamed into ``\texttt{coinhive.html}'' and modified, yet the original
mining functionality was intact.

Overall, these observations favor the intuition that, given the small hash rate for smartphone-based
mining~\cite{saad2018end}, the malicious actors would not spend resources on implementing the mining functionality from
scratch, but rather use the libraries that are already available.  

By looking at the used third-party libraries and the code of the miners from our sample, we were able to determine that
\textbf{586} miners target the Monero cryptocurrency, \textbf{5} miners target Ethereum, \textbf{5} miners target
Litecoin, and \textbf{3} miners have been created to test the capability of mobile devices for mining Bitcoin.
\textbf{91} binary miners in our sample rely on third-party mining libraries that can be used to mine multiple
cryptocurrencies (Monero, Litecoin, Ethereum, Bitcoin, and others).

\begin{figure}[!ht]
    \centering
    \includegraphics[width=.6\columnwidth]{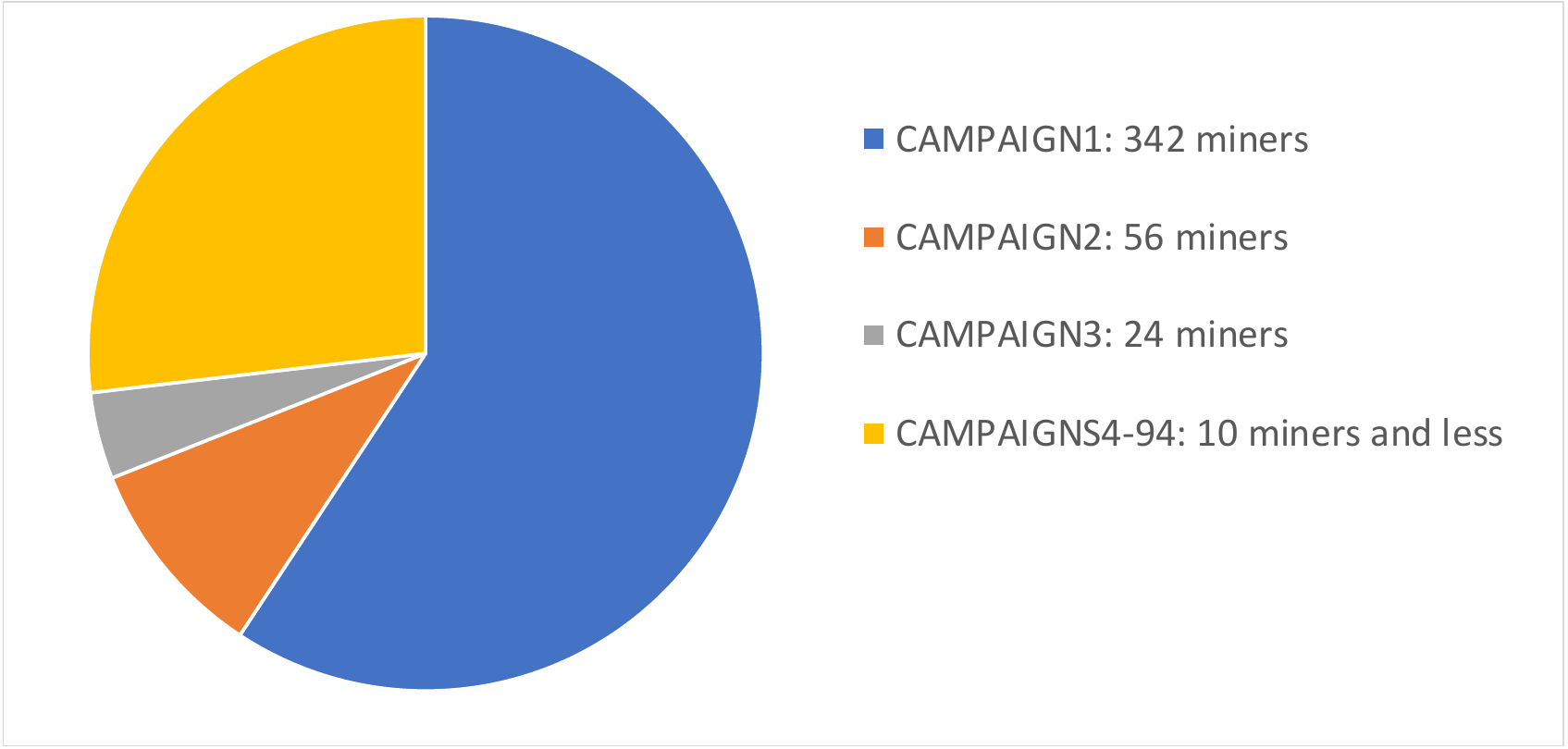}
    \caption{The sizes of mining campaigns}
    \label{fig:campaign-stats}
\end{figure}

\paragraph{Mining campaigns.} Similarly to previous works Konoth et al.~\cite{konoth2018minesweeper} and Hong et
al.~\cite{hong2018you}, we tried to identify the \emph{mining campaigns} by grouping the sets of miners that are likely
to share the same origin, and therefore may share the benefits from the mined cryptocurrency. As most of the miners from
our sample reuse the same third-party mining libraries, it is difficult to identify the same origin by looking at the
similar code patterns in the apps. Therefore, we assume that two miners belong to the same campaign only if they share
the same mining credentials used for authentication with the mining services (e.g., the cryptocurrency \emph{wallet id}
and/or CoinHive \emph{site key}).

Figure \ref{fig:campaign-stats} shows the distribution of the sizes of the mining campaigns found within our sample.
Overall, we found \textbf{94} unique mining campaigns, with the largest campaign enclosing \textbf{342} miners, two
smaller campaigns enclosing \textbf{56} and \textbf{24} miners respectively, and \textbf{91} small campaigns of \textbf{10}
miners and less. The rest of the apps are \emph{benign} miners that did not contain any mining credentials, or
\emph{illicit} miners for which we could not retrieve these credentials. Therefore, we could not consider such miners to
be a part of a mining campaign. 

At this stage we cannot conclude whether the small mining campaigns that we found are indeed small ``in the wild'', as
this requires further large-scale data collection and analysis. However, the two relatively large campaigns suggest that
in the wild there may be many Android apps created or, more likely, repackaged by the same malicious developer that is
actively trying to maximize her mining profit. Indeed, it is relatively inexpensive to repackage an already existing app
and insert only the mining code. We have seen many examples that support this conclusion, see Section
\ref{sec:detection_android_miners}.

When searching for the mining credentials used in the biggest mining campaign that we identified, we found that a
security researcher has already reported\footnote{\url{https://twitter.com/fs0c131y/status/950082654891802630}} this
mining campaign. Still, our sample contains more apps than it was originally reported (\textbf{342} versus 291).
Moreover, at least \textbf{24} of apps from our sample that belong to this campaign do not share similar code (unlike
the apps seen by the researcher), suggesting that the campaign might be even bigger in the wild.
In our sample there are another \textbf{64} miners that correspond to \textbf{17} mining campaigns which mining
credentials have been reported in whitepapers and blogposts by other researchers. Yet, we have collected \textbf{173}
miners that correspond to \textbf{76} campaigns that have not been previously reported. In particular, the second
largest \emph{illicit} campaign shown on Figure \ref{fig:campaign-stats} is has not been reported before.

\paragraph{Miner-related apps.}
We have also encountered \textbf{34} Android apps that we refer to as \emph{miner-related}. While these apps do not
perform any cryptocurrency mining, they are riddled with keywords, links, and mining credentials relevant to the real
mining apps.  Such apps may pose additional challenges for automated miner detection approaches, and it is therefore
important to consider their presence ``in the wild''. We include these apps as a separate category in the dataset we
have built, because they are valuable confusing data points. Below we briefly describe them.

\textbf{12} of these apps have useful functionality: e.g., they either monitor the value of cryptocurrencies, or serve
as cryptocurrency wallets, or simply ask for donations in cryptocurrencies (for apps of the latter case we found a match
for a cryptocurrency wallet). These applications can serve as confusion points since various static indicators would
suggest the presence of mining code (see Section \ref{subsec:static-detection}).
The rest of \textbf{22} miner-related apps are \emph{scam}. They do not have any useful functionality, and claim to be
legitimate mining apps. Their monetization comes from either showing paid ads, or tricking their users into paying for
an ``upgraded'' version of the app: for example, the ``basic'' version may claim that it does not support transferring
the mined funds, until a sufficient amount of a cryptocurrency is mined. In particular, 2 of these apps employ a trick
to improve their ratings: from the start they promise the user 50,000 Satoshis (0.0005 Bitcoins) for ``free'' if the
user rates the app. Such apps correspond to another possible source of confusion for automated detection approaches:
while, an app claims it is a cryptocurrency miner and should be immediately considered as a positive data point (e.g.,
by classification approaches from the area of machine learning), they neither contain the mining code, nor manifest the
runtime behavior typical to the mining apps (see Section \ref{subsec:dynamic_detection}).

\begin{table}[!ht]
    \begin{minipage}{.5\linewidth}
		\scriptsize
		\centering
		\begin{tabular}{lr}
			\toprule
			\textbf{Android permission}                     & \textbf{\# Apps (\%)} \\
			\midrule
			android.permission.INTERNET                    & 34 (100.00\%)   \\
			android.permission.ACCESS\_NETWORK\_STATE       & 34 (100.00\%)    \\
			android.permission.WAKE\_LOCK                   & 22 (64.70\%)    \\
			com.google.android.c2dm.permission.RECEIVE    &  19 (55.88\%)    \\
			android.permission.ACCESS\_WIFI\_STATE     & 15 (44.11\%)    \\
			android.permission.WRITE\_EXTERNAL\_STORAGE      & 13 (38.23\%)    \\
			android.permission.CAMERA                     & 12 (35.29\%)    \\
			android.permission.RECEIVE\_BOOT\_COMPLETED      & 11 (32.35\%)    \\
			android.permission.VIBRATE                       & 9 (26.47\%)    \\
			android.permission.ACCESS\_COARSE\_LOCATION        & 8 (23.52\%)    \\
			\bottomrule
		\end{tabular}
		\caption{Top 10 permissions in miner-related apps}
		\label{tab:top-perms-miner-rel}
    \end{minipage}%
    \begin{minipage}{.5\linewidth}
		\scriptsize
		\centering
		\begin{tabular}{lr}
			\toprule
			\textbf{Android system event}                                & \textbf{\# Apps (\%)}\\
			\midrule
			com.android.vending.INSTALL\_REFERRER                          & 19 (55.88\%)    \\
			android.intent.action.BOOT\_COMPLETED                         & 16 (47.05\%)    \\
			com.google.android.c2dm.intent.RECEIVE                       & 16 (47.05\%)    \\
			android.net.conn.CONNECTIVITY\_CHANGE                          & 7 (20.58\%)    \\
			android.intent.action.QUICKBOOT\_POWERON                    & 5 (14.70\%)    \\
			android.intent.action.PACKAGE\_ADDED                       & 5  (14.70\%)    \\
			android.intent.action.ACTION\_POWER\_CONNECTED                         & 4  (11.76\%)    \\
			android.intent.action.ACTION\_POWER\_DISCONNECTED               & 4  (11.76\%)    \\
			android.intent.action.MY\_PACKAGE\_REPLACED                  & 4  (11.76\%)    \\
			android.net.wifi.WIFI\_STATE\_CHANGED                         & 4  (11.76\%)    \\

			\bottomrule
		\end{tabular}
		\caption{Top 10 system events used by miner-related apps}
		\label{tab:top-actions-miner-rel}
    \end{minipage} 
\end{table}

Table~\ref{tab:top-perms-miner-rel} summarizes top 10 permissions requested by miner-related applications, and Table~\ref{tab:top-actions-miner-rel} shows top 10 system events that miner-related apps subscribe to. The requested permissions and monitored system events in these apps largely coincide with the statistics reported for the mining sample in Tables~\ref{tab:top-perms}\&\ref{tab:top-actions}. This further shows that lightweight, keyword-based detection approaches for miners may produce many false-positives. At the same time, permissions and system events may indicate that miner-related apps, especially of the \emph{scam} type, are supplied by malicious actors. Yet, our sample of miner-related applications in the dataset is relatively small; thus a larger investigation of a larger population of such applications is warranted.

\subsection{Examples of Mining Applications}

\begin{figure}[!ht]
    \centering
    \begin{subfigure}[b]{0.3\textwidth}
        \includegraphics[width=\textwidth]{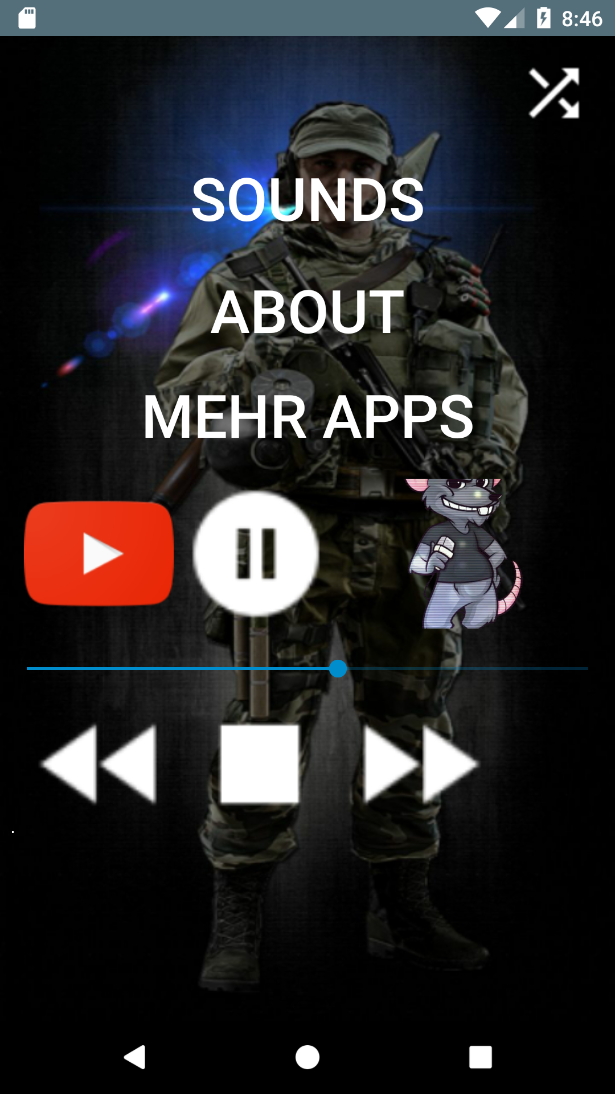}
        \caption{Illicit miner}
        \label{fig:illicit_miner_screenshot}
    \end{subfigure}
    \qquad
    \begin{subfigure}[b]{0.3\textwidth}
        \includegraphics[width=\textwidth]{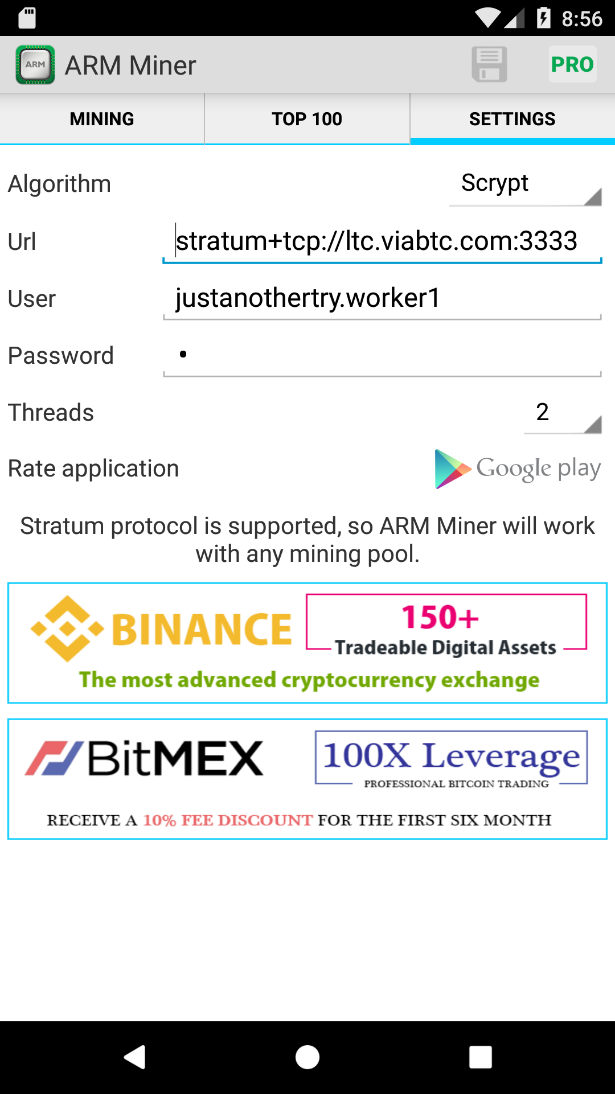}
        \caption{Legal miner}
        \label{fig:legal_miner_screenshot}
    \end{subfigure}
    \caption{Screenshots of miner apps}
    \label{fig:miners_screenshot}
\end{figure}

The screens in Figures~\ref{fig:illicit_miner_screenshot} and \ref{fig:legal_miner_screenshot} demonstrate examples of two mining applications. The first app is an illicit miner\footnote{SHA256 \texttt{a3f376a5c74e1fe112786b4ad450a6b3976226e2164b106653483522adf6bced}}. It looks like an app that was created just for fun and provides very basic functionality playing a funny song. However, invisibly it mines the Monero cryptocurrency in a hidden Web browser. The second example is a legal miner created specifically for mining Bitcoin on ARM devices\footnote{SHA256 \texttt{727cd092ed478453c2f19d180e1aa8fd22e43dc9cf24772c5ae2ca36cf9dbc4e}}. In this miner users need to configure their own mining credentials and run the miner. It is a binary miner that exploits a standalone executable \texttt{minerd}. Both apps have been previously hosted on Google Play.

\subsection{VirusTotal Analysis Results}%
\label{subsec:vt_dataset_description}

We checked each sample from our mining dataset using the VirusTotal service. Using their API, we downloaded the VirusTotal extended analysis reports for each app in our dataset, obtaining the latest report version for the time of writing. If a report was not found, i.e., a sample had not been uploaded to VirusTotal before, we submitted the application on our own. 

We were the first who found and uploaded at least 5 samples to VirusTotal\footnote{We have not collected this statistics from the start, therefore, we can confirm only 5 cases.}. Among previously seen samples, an app from our dataset was checked by VirusTotal at the earliest in October 2013, while the most recent one was uploaded in March, 2019. 

\begin{figure}[!ht]
    \centering
    \includegraphics[width=0.6\columnwidth]{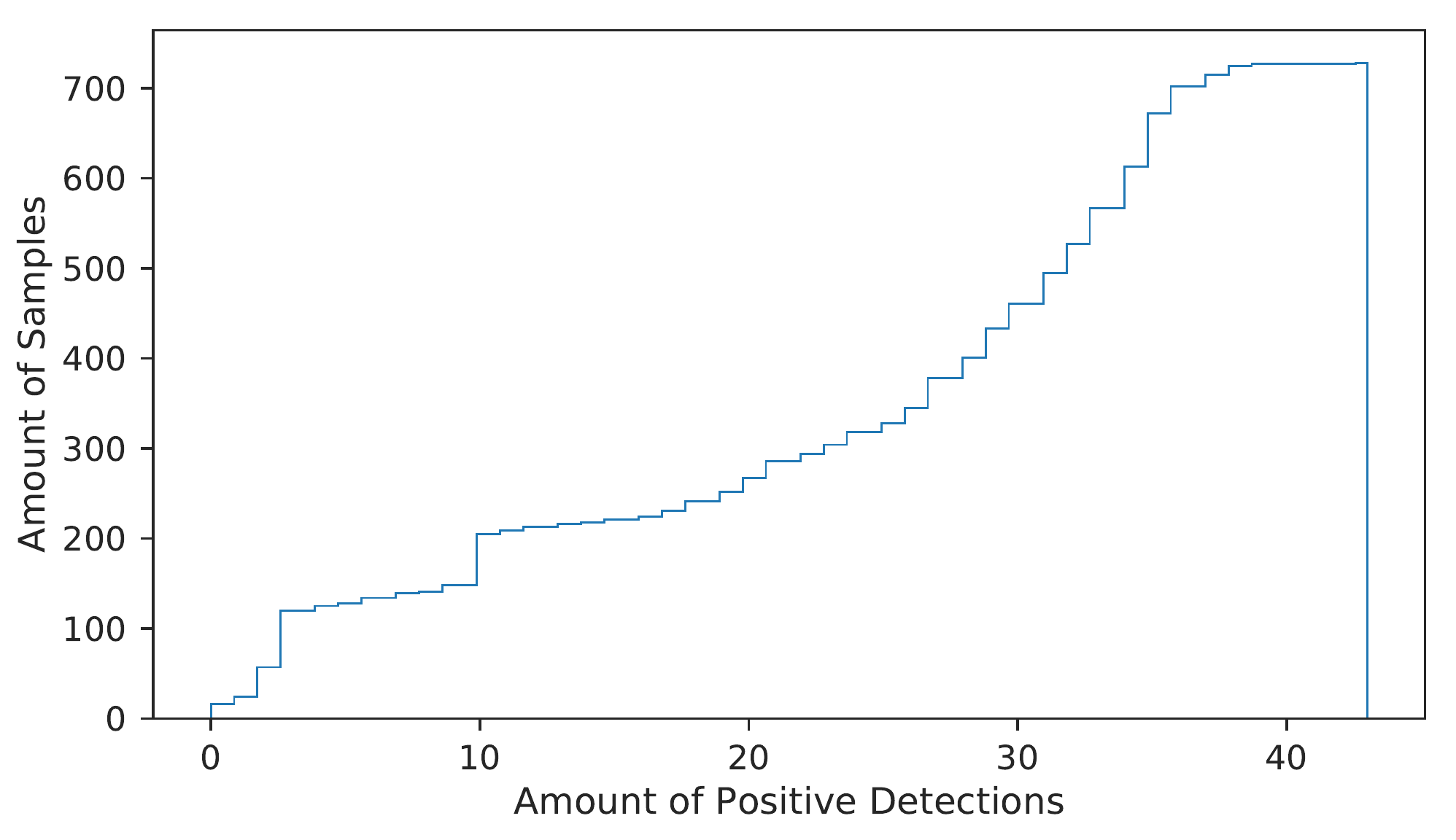}
    \caption{CDF of positive detection by VirusTotal scanners}
    \label{fig:detections_cdf}
\end{figure}

All applications from our dataset have been checked by at least 1 out of \textit{77 antivirus products} aggregated on the platform. Figure~\ref{fig:detections_cdf} shows the Cumulative Distribution Function (CDF) representing the amount of antivirus scanners that detected each application from our dataset. On average, a sample in our list is marked as malicious by \textit{22 scanner}. Maximum, a sample in our dataset is detected by \textit{43 different scanners}. This shows that even old, well-known samples are not recognized by all scanners.

In Figure~\ref{fig:detections_cdf}, we can spot three high steps at 3, 10 and 35 detections (63, 57 and 59 new samples correspondingly). These steps could have appeared because some scanners have similar detection engines, or they share antivirus databases. Indeed, Table~\ref{tab:correlated_scanners} proves this assumption. It shows that the results of some scanner pairs are either identical or highly correlated. 

\begin{table}[!ht]
\centering
\scriptsize
\begin{tabular}{llr}
\toprule
\textbf{Scanner 1}  &  \textbf{Scanner 2}  &  \textbf{Correlation} \\
\midrule
    AntiVir     & Commtouch   &  1.000 \\
                & ByteHero    &  1.000 \\
    ByteHero    & Commtouch   &  1.000 \\
    Agnitum     & Commtouch   &  1.000 \\
    Commtouch   & Norman      &  1.000 \\
    ByteHero    & Norman      &  1.000 \\
    TACHYON     & nProtect    &  1.000 \\
    Agnitum     & Norman      &  1.000 \\
                & AntiVir     &  1.000 \\
    AntiVir     & Norman      &  1.000 \\
    Agnitum     & ByteHero    &  1.000 \\
    AVG         & Avast       &  0.997 \\
    Kaspersky   & ZoneAlarm   &  0.990 \\
    BitDefender & Emsisoft    &  0.984 \\
                & GData       &  0.927 \\
    Emsisoft    & GData       &  0.922 \\
    BitDefender & MAX         &  0.894 \\
    Emsisoft    & MAX         &  0.886 \\
    GData       & MAX         &  0.862 \\
    Arcabit     & BitDefender &  0.857 \\
\bottomrule
\end{tabular}
\caption{Top 20 highly correlated scanner pairs}%
\label{tab:correlated_scanners}
\end{table}

Interestingly, \textbf{16} miners are not detected by any antivirus product. Table~\ref{tab:undetected_samples} reports
SHA256 hashes of these miners and the accompanied data that we extracted from VirusTotal reports. We determined
\textbf{10} out of these 16 apps as \emph{legitimate} miners, and \textbf{6} of them as \emph{illicit}. The miners from
this table are not new: the oldest is dated back to 2013. However, even the illicit miners among these apps are still
not recognized as malicious or unwanted. For \textbf{4} apps this can be explained by the fact that the mining script is
stored in the encrypted form and is being decrypted at runtime, and \textbf{2} of the undetected apps use obfuscation.
Based on these results, we cannot not draw firm conclusions whether mining functionality is deemed malicious by
VirusTotal scanners, as \textbf{10} legitimate miners are also detected as malware. It could be also that our miners contain
some other malicious payloads, even though our analysis have not revealed such evidence for legitimate miners (but there
were ``also-malicious'' illicit miners).

\begin{table*}[!ht]
\centering
\scriptsize
\begin{tabular}{lcllrrrr}
\toprule
    \textbf{SHA256} & \textbf{Illicit} & \textbf{First seen} & \textbf{Last seen} &   \textbf{Amount of} &      \textbf{Amount of} & \\ 
                    &                  &       \textbf{date} &      \textbf{date} & \textbf{submissions} & \textbf{unique sources} & \\ 
    \midrule
    aa200375c8422f3e034b122aa45e59a289b6c356b2301c4651189c27a895d9b0 & \ding{54} &  2013-10-13  &  2015-03-14  &  4 &  2 &  \\ 
    76ae303c82d8233414694ff803c2a22bd82dc1ff1bab1341f9932a238b6b0efc & \ding{54} &  2017-07-28  &  2017-07-28  &  1 &  1 &  \\ 
    f8f936810980d14ab41abb91d4fb0bba32c083e6846623d4320ea45053e8ea6d & \ding{54} &  2017-09-27  &  2017-09-27  &  1 &  1 &  \\ 
    d735cf3732d00ce43d1a36bc77123770d5d611f09d39b8576e3698a9a2ebda87 & \ding{54} &  2017-12-01  &  2017-12-01  &  1 &  1 &  \\ 
    c491cbabb604a59c99e5be0e1808f43e1d21b94be524c4d1c759c8bbbd452509 & \ding{52} &  2018-01-09  &  2018-01-09  &  1 &  1 &  \\ 
    609941fdf62a6f9d186a7714bd4238e0d2c531badd96a69dbb2dce2b4f1d5248 & \ding{54} &  2018-02-28  &  2018-07-18  &  8 &  6 &  \\ 
    be11f2929b4383f1bcf020c8d7d8b4ef0172c5c5a4e468271ebb87f4b14db876 & \ding{54} &  2018-03-02  &  2019-04-24  &  4 &  3 &  \\ 
    c471ca1989d7fc7662ea3ba5bf0bcc79d8790fe4770acaaabd10dafadb7ee362 & \ding{54} &  2018-05-03  &  2018-05-03  &  2 &  2 &  \\ 
    630cf7f1728e8a592aa016171be1f7852f70baaed5267398417f8d91b9d14acb & \ding{54} &  2018-05-29  &  2018-09-08  &  2 &  2 &  \\ 
    f61e31ee2f27f2815e7720cad5920b750d10e01788cd78f7fbd81ba1c31dfeb3 & \ding{52} &  2018-07-13  &  2018-07-13  &  1 &  1 &  \\ 
    7acb35a690d02a34a404cae9ccd3f9b25558e43fd143514c7b42f225aa3663a3 & \ding{54} &  2018-08-05  &  2018-10-13  &  2 &  2 &  \\ 
    17b56ef3a43c6cd4245113ada9a9ff0364754fc6947d05e9f9acb8e6630f9d27 & \ding{54} &  2018-08-23  &  2018-08-23  &  1 &  1 &  \\ 
    2e5ba00cc3caa0a4801f2b0580829cee0577e4b05719e86ea7e5690c961d5dae & \ding{52} &  2019-01-30  &  2019-01-30  &  1 &  1 &  \\ 
    33db4abf2526b4bda22559e41052ea12362c25e96fd0ebd49becd470694e57de & \ding{52} &  2019-02-02  &  2019-04-05  &  2 &  2 &  \\ 
    cb6546a785af3aa2dfee434e25e66ca8691e56909a64e3e317a91fb4e2d5bd1b & \ding{52} &  2019-02-10  &  2019-02-10  &  1 &  1 &  \\ 
    ec8433cd5a06aaafe251361ec304dbc438272a5fef49cf7c5c45a63caecff375 & \ding{52} &  2019-03-02  &  2019-03-02  &  1 &  1 &  \\ 
\bottomrule
\end{tabular}
\caption{Samples not detected by the VirusTotal scanners}%
\label{tab:undetected_samples}
\end{table*}

\begin{figure}[!ht]
    \centering
    \begin{subfigure}[b]{.49\textwidth}
        \centering
        \includegraphics[width=\columnwidth]{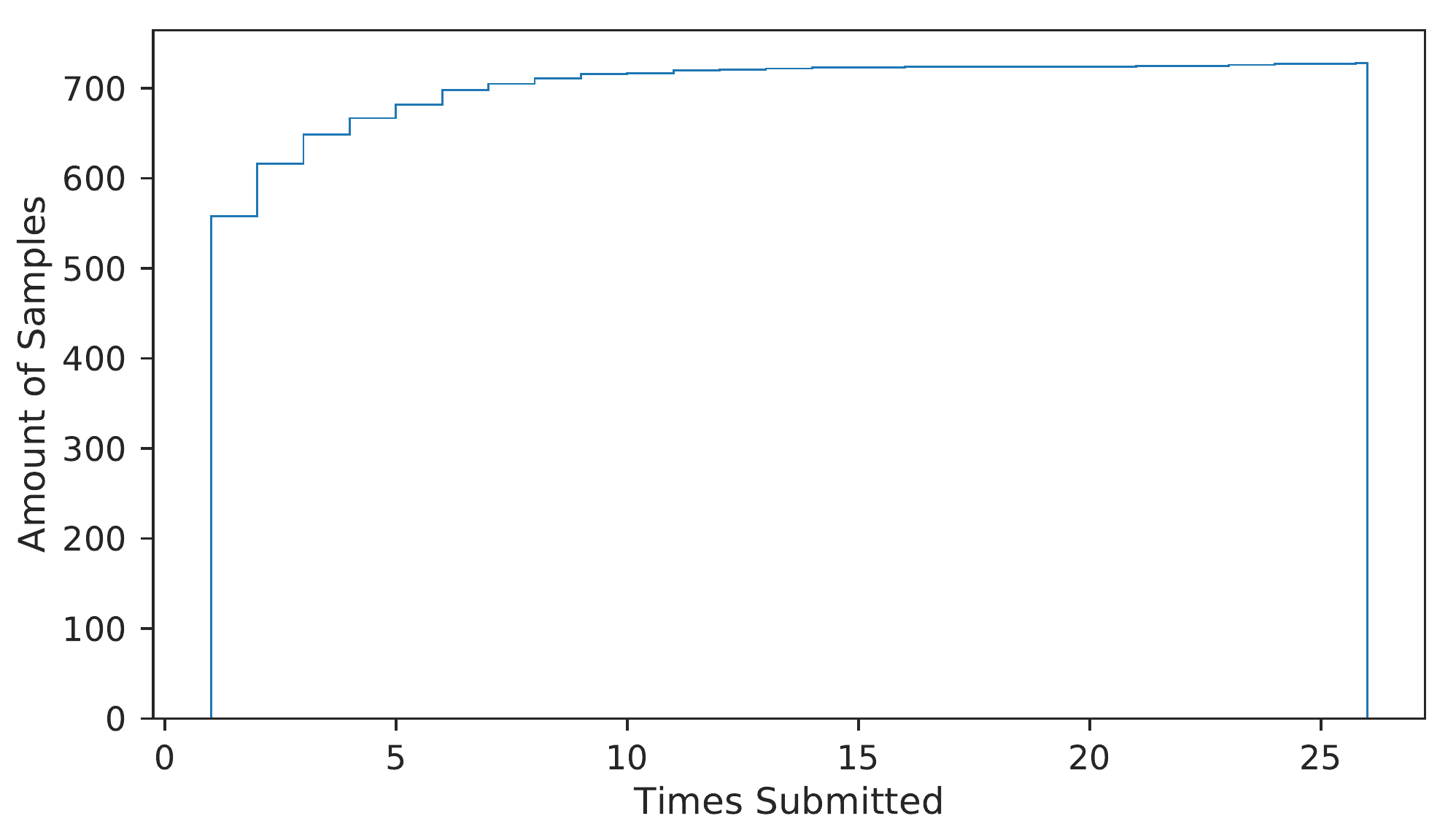}
        \caption{CDF of sample submissions}
        \label{fig:times_submitted_cdf}
    \end{subfigure}
    \begin{subfigure}[b]{.49\textwidth}
        \centering
        \includegraphics[width=\columnwidth]{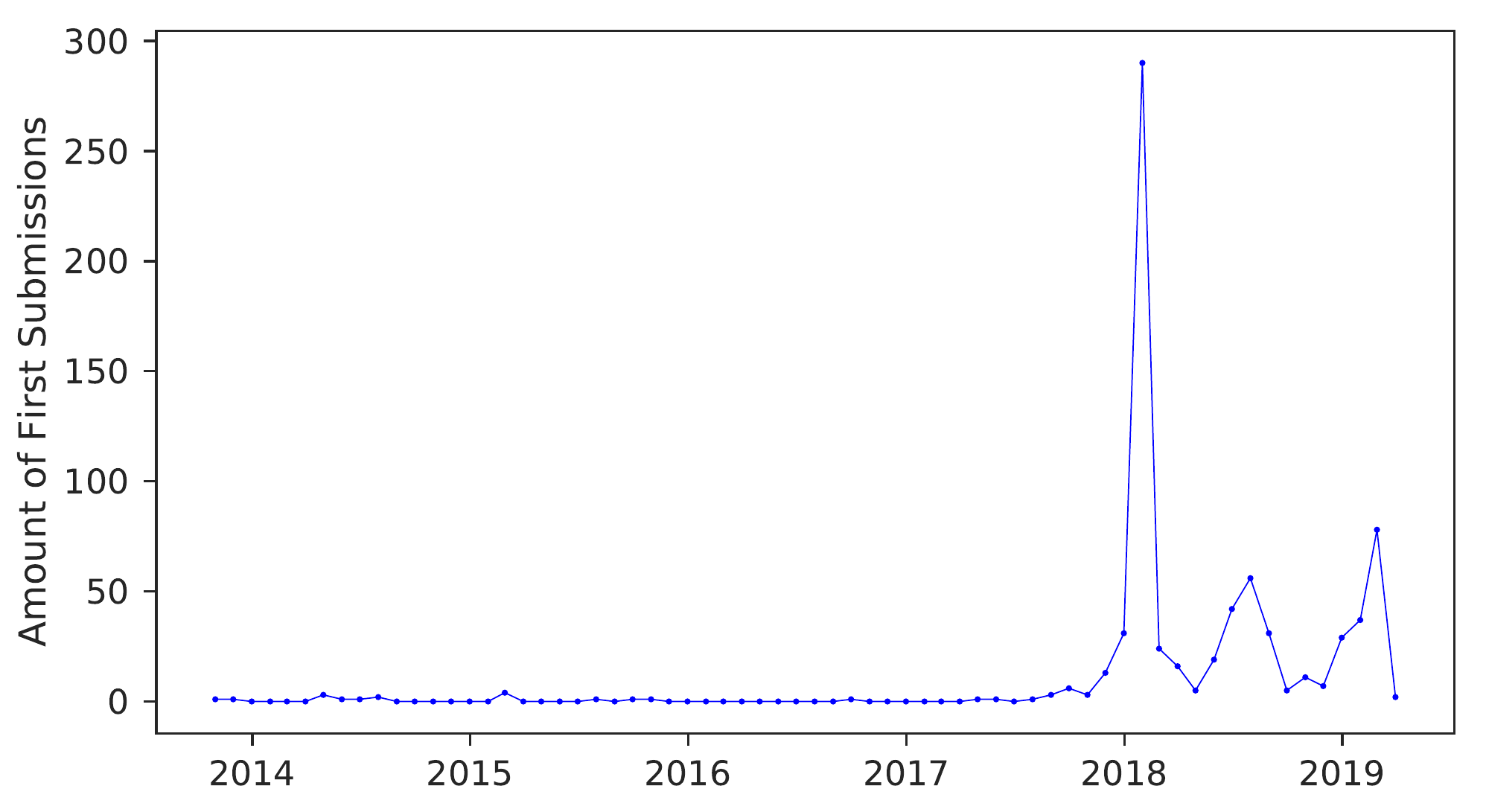}
        \caption{Distribution of new submissions to VirusTotal}
        \label{fig:first_seen_monthly_aggregated}
    \end{subfigure}
    \caption{Sample submissions to \emph{VirusTotal}}
    \label{fig:cdf_example}
\end{figure}

Figure~\ref{fig:times_submitted_cdf} shows how many times a sample from our dataset has been submitted to VirusTotal. On average, this value is around \textit{1.85}. Indeed, 558 samples have been uploaded to VirusTotal only once, while the most frequently submitted sample has been uploaded 26 times. 

Using the information on the date when a sample was submitted to VirusTotal first time, we evaluated if our dataset is relatively new. To achieve this goal, we aggregated the samples by months when they have been first spotted on VirusTotal. Figure~\ref{fig:first_seen_monthly_aggregated} shows the timeline when the apps from our list were submitted to VirusTotal for the first time. Several interesting observations can be derived from this figure. First, we can see that our dataset is relatively new: the majority of applications have been first spotted on VirusTotal in 2018. Second, the figure shows that before the last quarter of 2017 there were a few submissions of new miners, while in the beginning of 2018 we observe huge spike. This phenomenon can be explained by cryptocurrencies popularity in general. Indeed, before 2017 the interest to the cryptocurrencies was mostly driven by niche experts and geeks. However, in 2017 the price of Bitcoin and other cryptocurrencies started to grow exponentially. This attracted the attention of malware developers, who started to explore this market. Clearly, a miner is a very attractive type of malicious application because it directly earns money for the developer, while almost no efforts need to be spent on its preparation and distribution through repackaging. Not surprisingly, in the end of 2017 antivirus companies started to consider Android miners as harmful applications~\cite{CoinMiners_ZDNet,cyberthreatreport2018}.           

We have also ranked the VirusToal scanners according to their ability to detect mining applications on our the dataset of manually confirmed miners and their detection results. We assigned +1 point to each true positive and -1 point to each false negative; if a scanner failed to scan a sample or VirusTotal does not have the data, we gave 0 points. The final score is calculated as sum of these points. Table~\ref{tab:scanners_rating} reports the top 10 VirusTotal scanners based on this score. 

Note that Table~\ref{tab:scanners_rating} shows the rating based only on our dataset consisting on the samples from one class (miners).  This could introduce a deviation in our ranking because a scanner that marks all submitted files as malicious would take the first place in our ranking. To make a more fair list, we would need to get also the list of benign applications and test the scanners on them. However, a comprehensive evaluation of the VirusTotal scanners is out of our scope for this work. 

\begin{table}[!ht]
\centering
\scriptsize
\begin{tabular}{lrrrr}
    \toprule
    \textbf{Scanner} & \textbf{Final} &       \textbf{True} &      \textbf{False} &  \textbf{Failed /} \\
                     & \textbf{score} &  \textbf{positives} &  \textbf{negatives} &   \textbf{no data} \\
    \midrule
                     Sophos &  514 &  621 &  107 &    0 \\
              CAT-QuickHeal &  478 &  603 &  125 &    0 \\
                      DrWeb &  474 &  601 &  127 &    0 \\
                 ESET-NOD32 &  394 &  561 &  167 &    0 \\
                     Ikarus &  346 &  512 &  166 &   50 \\
                      Avira &  285 &  505 &  220 &    3 \\
                     McAfee &  266 &  497 &  231 &    0 \\
      SymantecMobileInsight &  256 &  408 &  152 &  168 \\
                  ZoneAlarm &  254 &  489 &  235 &    4 \\
                  Kaspersky &  252 &  489 &  237 &    2 \\
     \bottomrule
\end{tabular}
\caption{Top 10 VirusTotal scanners evaluated on our dataset}%
\label{tab:scanners_rating}
\end{table}



\section{Detecting Android Miners}%
\label{sec:detection_android_miners}


\subsection{Static Indicators}
\label{subsec:static-detection}

In this Section we describe the heuristics that we identified and used when performing manual analysis of potential
miner apps (Section \ref{sec:dataset}). These heuristics can be used as static indicators for pinpointing potential
Android miners across large amounts of apps.

\paragraph{Static heuristics.}
As we describe in Section \ref{sec:sample_statistics}, the vast majority of our miners use third-party mining libraries,
and often the code of these libraries is used without any changes. Therefore, finding the presence of the code of these
libraries will indicate a potential miner with high degree of certainty. For libraries written in JavaScript we take
note of the distinctive code patterns and strings.  For libraries written in Java (e.g., \emph{CoinHive Android SDK}
shown in Table \ref{tab:libraries}) we take note of distinctive components of the library such as package names,
classes, and smali code patterns. For native libraries and executables we take note of their filename and SHA256 hash
code, as well as specific string patterns that can be present inside them. For instance, most of the native mining
libraries listed in Table \ref{tab:libraries} had a distinctive help menu that lists the available mining parameters and
settings. When we see an unknown binary file that could be a mining library we can obtain its hexdump using command line
tools such as \emph{xxd} and compare string patterns inside the binary file against the string patterns retrieved from
known mining libraries.
While such approach cannot beat sophisticated obfuscation techniques, it may be still helpful to uncover a large set of
miners where the library (or its parts) is used as is. We find that this simple heuristic is quite powerful, allowing us
to find many miners that were difficult to spot otherwise. 

We illustrate this heuristic with the \emph{CoinHive Android SDK} library (Table \ref{tab:libraries}).  The library
implements a convenient Java wrapper around the \emph{CoinHive} JavaScript API. Thus, the library can be added directly
to an Android project as a dependency, and a CoinHive miner instance can be created and launched from within the Java
code, as shown in Listing~\ref{lst:coinhive-sdk-init}. The \texttt{CoinHive} Java class contains JavaScript-to-Java
bindings to the file called ``\texttt{engine.html}'' located in the ``\texttt{resources/}'' folder of the SDK.  This
file includes the plain \emph{CoinHive API} JavaScript library (Table \ref{tab:libraries}).  Therefore,
\textit{javascript miners} that rely on this library can be relatively easily detected by searching for known code
patterns specific for the mining library (e.g., the smali code that corresponds to the miner initialization code shown
in Listing~\ref{lst:coinhive-sdk-init}), and/or for the code patterns present in the ``\texttt{engine.html}'' file.

\begin{envcode}{style=java, label={lst:coinhive-sdk-init}, caption={CoinHive Android SDK initialization example}}
public class App extends Application {
    @Override
    public void onCreate() {
        super.onCreate();

        CoinHive.getInstance()
                .init("YOUR-SITE-KEY") // mining credentials
                .setNumberOfThreads(4) // CPU threads
                .setThrottle(0.2)      // CPU throttle
    }
}
\end{envcode}

The miner initialization code for Web-based mining services, such as \emph{CoinHive API}, is typically inserted into
benign HTML/JavaScript resources of an app, and is loaded into an Android \texttt{WebView} UI element through the
``\texttt{WebView.loadUrl(...)}'' call -- this is quite similar to how the browser-based mining works in the
Web~\cite{konoth2018minesweeper} (the Android-specific code looks similar to the example we made on
Listing~\ref{lst:webview-example}).  In some cases, the JavaScript mining code is stored as a string constant inside the
\emph{smali} code and is passed directly into a \texttt{WebView} element. 

The authors of \emph{binary miners} typically place the mining libraries (e.g., ``\texttt{libcpuminer.so}'') or
standalone executables (e.g., ``\texttt{minerd}'' ELF executable) under the ``\texttt{res/raw/}'' or the
``\texttt{assets/}'' folder of an app archive.  These libraries are invoked either via the Android
``\texttt{System.loadLibrary(...)}'' interface, or by spawning separate application processes for executables (the
Android-specific code looks similar to the examples we made in Listing~\ref{lst:native-lib-example}
and~\ref{lst:executable-example}). 

We also looked for the mining credentials (e.g., cryptocurrency \emph{wallet} and \emph{site key} identifiers) passed
into the mining initialization code -- the presence of known mining credentials in apps immediately indicates that they
are most likely miners.  In general, we observed that \emph{illicit} miners contain the mining credentials somewhere in
the app code (\textbf{577} \emph{illicit} miners from our sample have hardcoded mining credentials). Therefore, to
significantly reduce the effort of quickly pinpointing new miners we built a collection of such identifiers.

We used several heuristics to retrieve the mining credentials. We observed that many \emph{javascript} and \emph{binary}
miners share similar miner initialization code patterns.  Therefore, after known third-party library code has been
located, it is easier to identify the code that initializes the mining and recover the mining credentials (this can also
help when the initialization code is obfuscated to a certain degree).  For example, in case of the \textit{CoinHive
Android SDK} library, we looked at the values of the parameters passed either to the \texttt{CoinHive} Java class
(Listing~\ref{lst:coinhive-sdk-init}), or the parameters passed directly into the ``\texttt{engine.html}''
file\footnote{E.g., ``\url{file:///android_asset/engine.html?site_key=...}''}. We also used regular expressions based
on the patterns of mining credentials for various cryptocurrencies. However, these regular expressions yielded too many
irrelevant strings: for instance, we often found strings like ``\emph{provideSHealthSyncedWorkoutsDAO}'', while we have
been searching for strings like ``\emph{NDMtBC8iLiUkEjUzKC8mYSQzMy4zYXxh}''. Therefore, we calculated the Shannon
Entropy metric~\cite{ma2010password} for such strings and only kept the strings for which this metric exceeded a certain
threshold (we empirically selected the value of 4.33). We checked the retrieved mining credentials and added them to our
string search, and found many more mining apps using this heuristic. 

Finally, to search through the apps for which we could not easily find known mining credentials or code patterns, we
used potential mining domains\footnote{We compiled a large list of known mining domains from various sources such as
\url{https://github.com/hoshsadiq/adblock-nocoin-list/blob/master/nocoin.txt}.} and simple keywords such as
\texttt{miner}, \texttt{bitcoin}, \texttt{stratum}, \texttt{monero}, \texttt{hashrate}, etc. While the keyword search
allows to find new previously unseen miners (which helped us a lot at the initial stages of our work), using it alone is
prone to large amounts of false-positives. For example, many non-miner apps that we encountered contain an adblock
functionality that actively tries to block known cryptocurrency mining domains (and thus, there will be a match to our
miner domain list).

\paragraph{Evasion techniques.}

We found cases when \emph{illicit} miners apply various evasion techniques and their combination to avoid detection:
the code fragments that initialize the mining process and contain the mining credentials may be not shipped with the
miner app itself, or this code can be encrypted within an app and only be decrypted at runtime. 

\begin{figure*}[!ht]
    \centering
    \includegraphics[width=0.8\columnwidth]{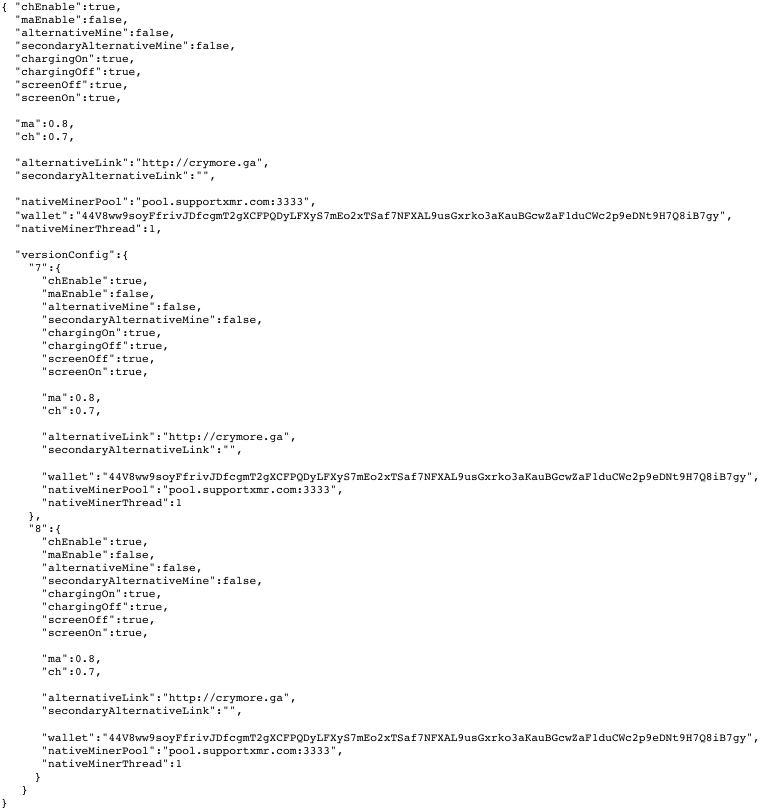}
    \caption{Illicit miner configuration served online}
    \label{fig:miner_setting}
\end{figure*}

For example, we found cases when an \emph{illicit} miner loads the mining credentials from a remote server upon the
application startup.  The download link is present within the code of the miner, but it has been obfuscated. However,
when we launched the app in the Android emulator, the \texttt{logcat} utility allowed us to see which link the app is
trying to connect to, and that it downloads a JSON file.  Upon further inspection of the file\footnote{The link is still
available at the time of writing: \url{https://raw.githubusercontent.com/cryptominesetting/setting/master/setting.txt}},
it became clear that it contained the mining credentials.  We provide a screenshot of this file in
Figure~\ref{fig:miner_setting}. In this figure, we can see the settings for the mining script, including the wallet
address, the preferred mining pool, and some configurations that allow to start mining when the device is charging and
not charging. By examining the \textbf{public GitHub
repository\footnote{\url{https://github.com/cryptominesetting/setting}}} where the link is hosted, we found several
other files that had similar structure but different mining wallets. We added these wallets to our miner-related
strings, however we have not found any apps that use them yet. This could be due to some other ways of hiding the mining
payload that we are not yet aware of, or possibly the owner of this repository is creating illicit cryptocurrency miners
for other application platforms (e.g., Google Chrome extensions).

Another approach for hiding miner credentials that we observed is as follows. The application resources contain an
\texttt{.html} resource file with the link to the CoinHive mining script, yet,  there seem to be no code that
initializes the mining process. Initially we thought that such apps had no mining capabilities. Yet, upon closer
inspection of other links embedded into the html pages, we identified a set of links to some JavaScript code located at
suspicious websites. Several of these links\footnote{The link is still available at the time of writing:
\url{https://api.kanke365.com/ads/app-tongyong-wk-7.js}} contained the code we have been looking for (shown in Listing
\ref{lst:remote-init-script}). Therefore, for improving the results of static Android miner detection, it is important
to download and inspect the remote resources, such as external links.

\begin{envcode}{style=javascript, label={lst:remote-init-script}, caption={Remote CoinHive initialization script example}}
var miner = new CoinHive.Anonymous('...');
miner.start();
\end{envcode}

We also found an interesting case when an \emph{illicit} miner consists of heavily obfuscated code (thus we initially
flagged it only as suspicious after a keyword match). The app actively tries to obtain the administrative permissions
from its users, and contains an encrypted file ``\texttt{assets/5a240bed02ae6}''. Upon thorough inspection of the app,
we we found the decryption key and were able to decrypt the file: we realized that the file contains a miner
initialization code which is being decrypted at runtime and dynamically called at via the Dalvik classloader (the main
reason why the app needs the admin privileges). 


We have also found that the majority of scam miners in our dataset are obfuscated, probably, to hinder inspection and to
make repackaging more difficult. Thus, it is necessary to perform runtime mining detection, not only in the cases, when
the mining code is not shipped within the Android app and/or is heavily obfuscated, but also to identify scam miners
that do not mine. To achieve dynamic detection, we have developed an approach described in the next section.

\subsection{Dynamic Detection}%
\label{subsec:dynamic_detection}
One of the most effective approaches to detect miners is to observe their dynamic behavior. Indeed, in order to gain maximum profit for the developers, a miner should use all available resources~\cite{saad2018end}. At the same time, in order to persist on the device, an illicit miner should conceal its mining activity, e.g., by applying throttling or doing this when the user does not use the phone. 

In this section, we propose an approach and a prototype called \tool\ that leverage machine learning for detecting the Android miners using dynamic features. In order to build this prototype we selected a dataset consisting of 200 Android applications: 100 miners and 100 benign apps.

For the miners dataset, we selected 100 apks from our sample that start the mining process immediately after they have launched. This is a
valid assumption because our tool is supposed to constantly monitor applications on a device and, thus, can detect the
moment when an app starts mining. Moreover, dormant miners do not cause damage for the user.

As the benign dataset, we randomly selected 100 apps from the local Google Play store among the ``Trending'', ``Top Apps'', and ``Top
Grossing'' application groups. These apps include various categories such as ``Gaming'', ``Education'',
``Sports'' and ``Shopping'', and their number of downloads ranged from 100 to more than 500M.

For each of these apps, we collected a set of traces that contain different dynamic parameter values generated by the corresponding application. We used the \emph{Snapdragon Profiler}~\cite{SnapdragonProfiler} to collect these traces. This is a tool developed by Qualcomm Technologies to profile execution of an Android app by collecting CPU, GPU, DSP, memory, power, thermal, and network data in order to find and fix performance issues. We ran each application from our dataset on LG Nexus 5 powered by the Snapdragon 800 system-on-chip running the Lineage OS 14.1 operating system (based on Android 7.1), and collected data about the low-level system events. Each application was exercised for 300 seconds. 

In a nutshell, each low-level system event is represented by 4 values: Process name, Timestamp, Metric name, and Metric Value. We consider as a \textit{metric timeseries} a sequence of the values with the corresponding timestamps that share the same application and profiled metric. For each application, we collected 15 different metrics: 1) Battery Current; 2) Battery Power; 3) CPU Branch Misses; 4) CPU Clock; 5) CPU Context Switches; 6) CPU Cycles; 7) CPU Cycles/Instruction; 8) CPU Instructions; 9) CPU Page Faults; 10) CPU Task Clock; 11) CPU Utilization Percent; 12) Memory Usage; 13) Rx Bytes (Total); 14) Tx Bytes (Total); and 15) Temperature. For each of the timeseries, we calculated 10 simple statistical values: 1) Minimum (Min); 2) Maximum (Max); 3) Average (Mean); 4) Median (Median); 5) Unbiased kurtosis (Kurt); 6) Unbiased skew (Skew); 7) Unbiased standard error of the mean (Sem); 8) Standard deviation (Std); 9) Mean absolute deviation (Mad); 10) Coefficient of variation (CV). Thus, for every application we obtained a feature vector consisting of 150 values. 

\begin{figure}[!ht]
    \centering
    \includegraphics[width=\textwidth]{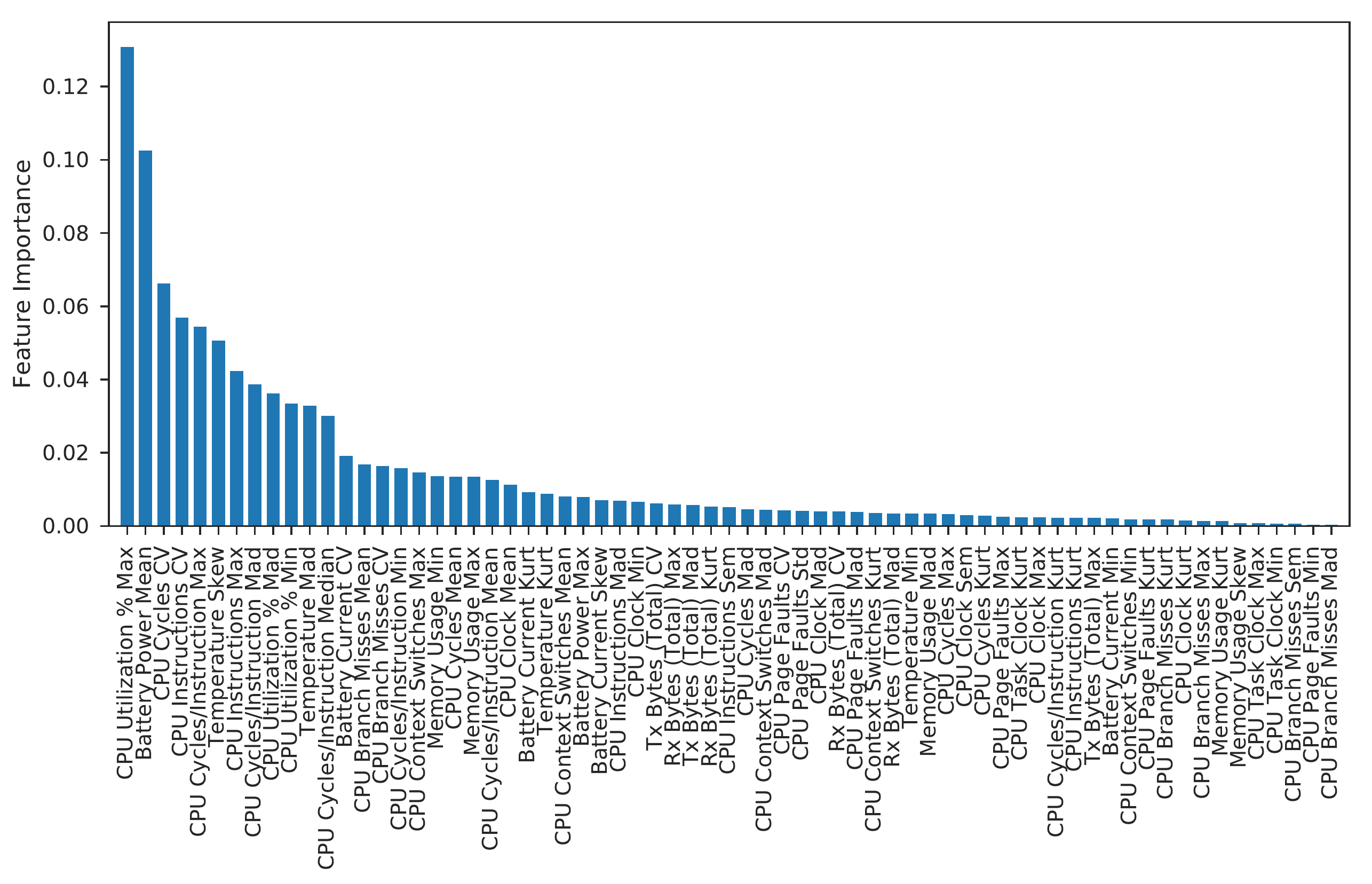}
    \caption{Feature importance}
    \label{fig:feature_importance}
\end{figure}

This amount of features is large, considering the size of our dataset. Therefore, we applied two feature selection techniques to eliminate excessive variables. It should be mentioned that we used the filtering techniques that perform cleaning only based on the internal properties of the dataset, without considering its connection to our application classes. First, we removed the features that have low variance in our dataset (threshold=0.1) using the scikit-learn library~\cite{scikit-learn}. This operation removed 20 features from our dataset. Second, we eliminated highly correlated features (Pearson correlation coefficient is more than 0.9). After this procedure, only 67 features were left to be used further (see Figure~\ref{fig:feature_importance} for the list). 

\begin{figure}[!ht]
    \centering
    \begin{subfigure}[b]{.49\textwidth}
        \centering
        \includegraphics[width=\textwidth]{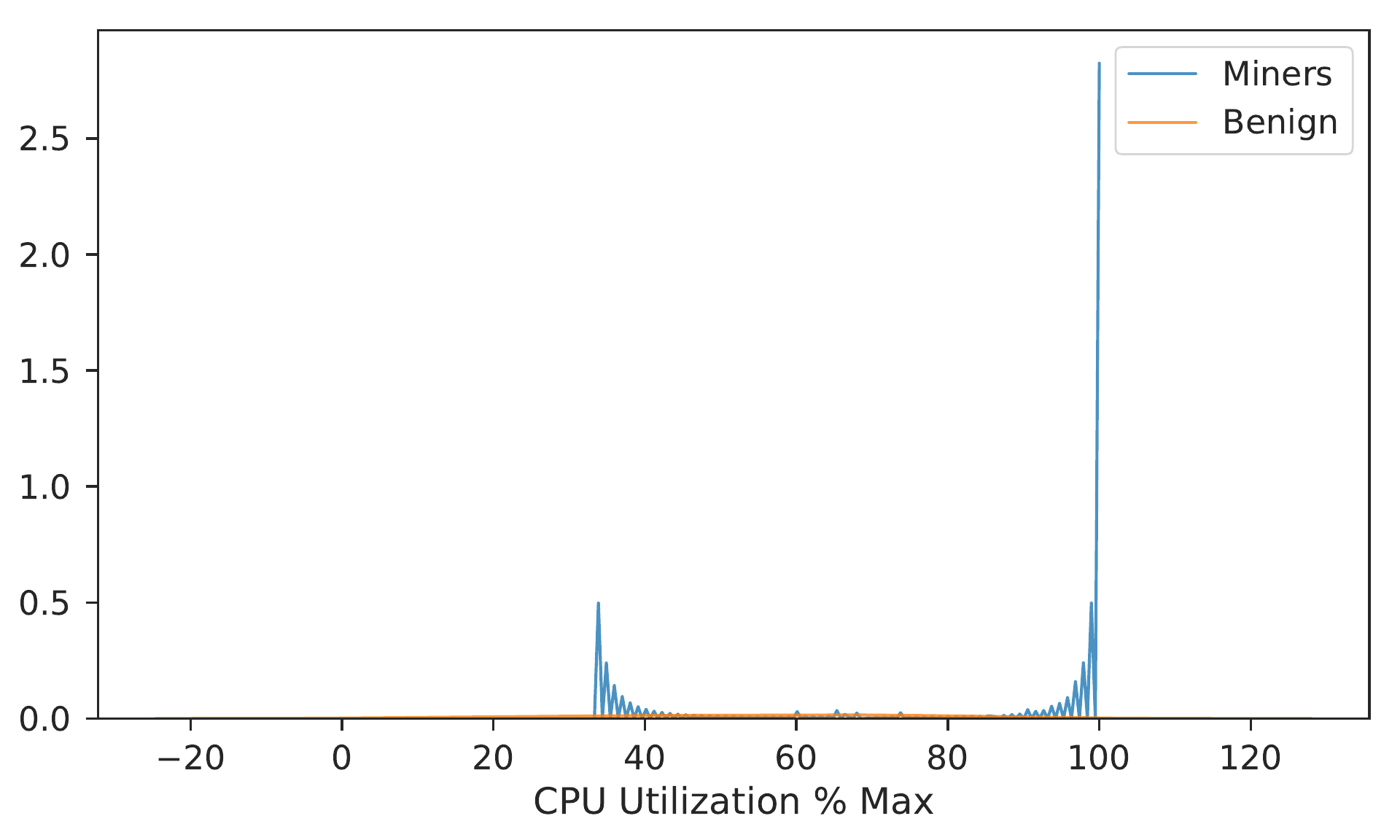}
        \caption{Maximum CPU Utilization \% KDE}
        \label{fig:cpu_utilization_perc_max_kde}
    \end{subfigure}
    \begin{subfigure}[b]{.49\textwidth}
        \centering
        \includegraphics[width=\textwidth]{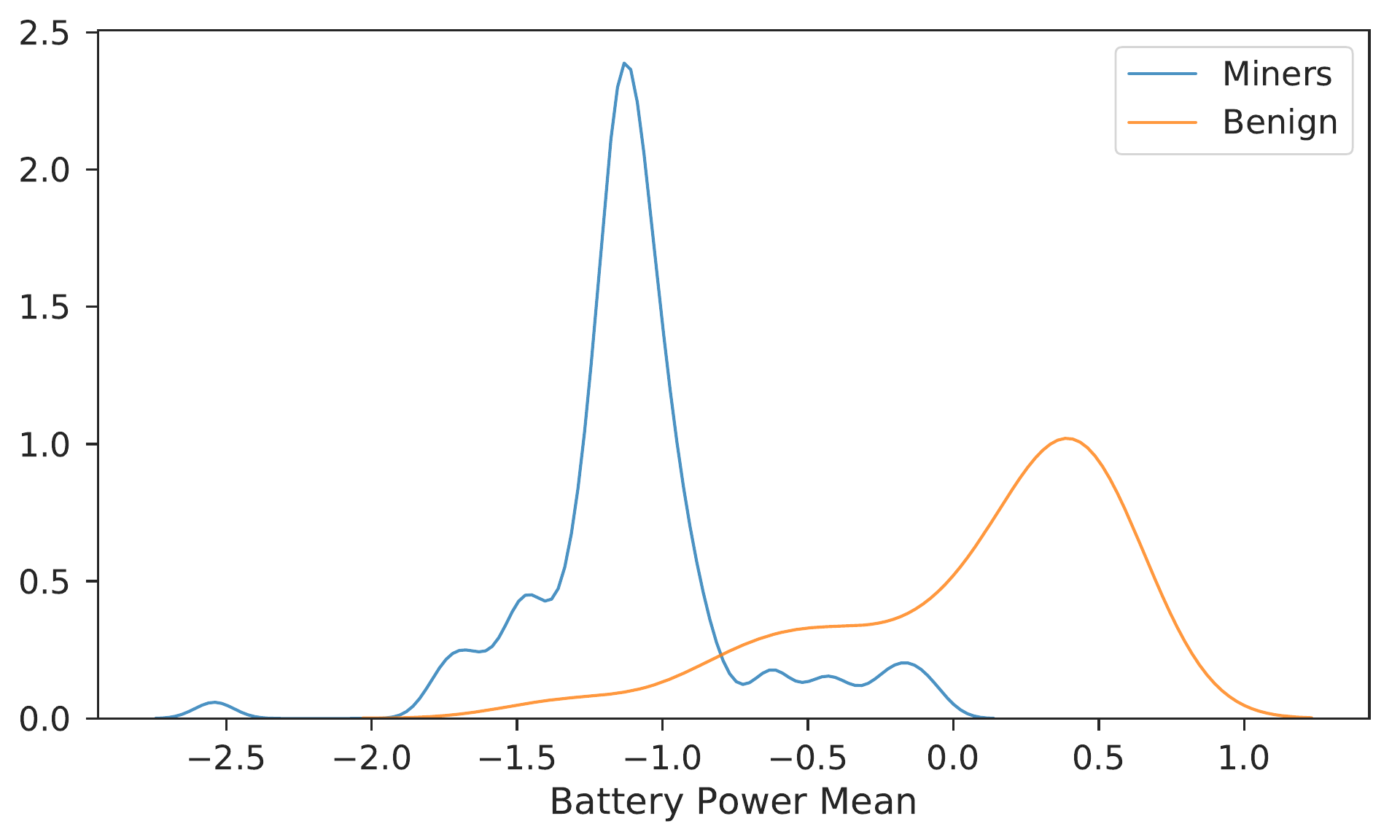}
        \caption{Battery Power Average KDE}
        \label{fig:battery_power_mean_kde}
    \end{subfigure}
    \caption{An example of dynamic features}
    \label{fig:example_dynamic_features}
\end{figure}

To detect the strongest features in our dataset\footnote{This information can be further used to collect only a subset of strong features on a device.}, we exploited the internal property of tree-based algorithms that calculate feature importances as a part of their training procedure. We trained a Random Forest classifier~\cite{scikit-learn} on our dataset. Figure~\ref{fig:feature_importance} lists the extracted features and shows their importance. The first two positions in this figure occupy the Maximum CPU Utilization \% and  Average Battery Power features. The corresponding Kernel Density Estimation (KDE) plots are shown in Figures~\ref{fig:cpu_utilization_perc_max_kde} and~\ref{fig:battery_power_mean_kde}. 

Several observations can be taken from these graphs. First, in Figure~\ref{fig:cpu_utilization_perc_max_kde}, a huge spike around 100\% could be observed for miners. This confirms that miners try to utilize all CPU resources on the device. At the same time, we also see some spikes around 30\% tick. This proves that some miners in our dataset throttled their mining capability, or used a subset of all available CPU cores. 

Second, the Maximum CPU Utilization \% KDE for benign applications is almost uniformly distributed along the $X$ axis. That means that there is no specific pattern of CPU utilization by benign apps. I.e., different applications consume on average different amount of CPU resources. The more intensive tasks a processor executes, the more power it requires. During our experiment, the phone collecting the dataset was attached to the computer through a USB cable. As a side-effect, during the dataset collection the phone was also charging. It can be seen on Figure~\ref{fig:battery_power_mean_kde} that when a benign application is executed the phone was actually charging (the extremum value is on the positive side). At the same time, the miners were consuming so much energy that the battery was even draining, even though the phone was attached to a source of energy. 

\begin{figure}[!ht]
    \centering
    \includegraphics[width=0.6\columnwidth]{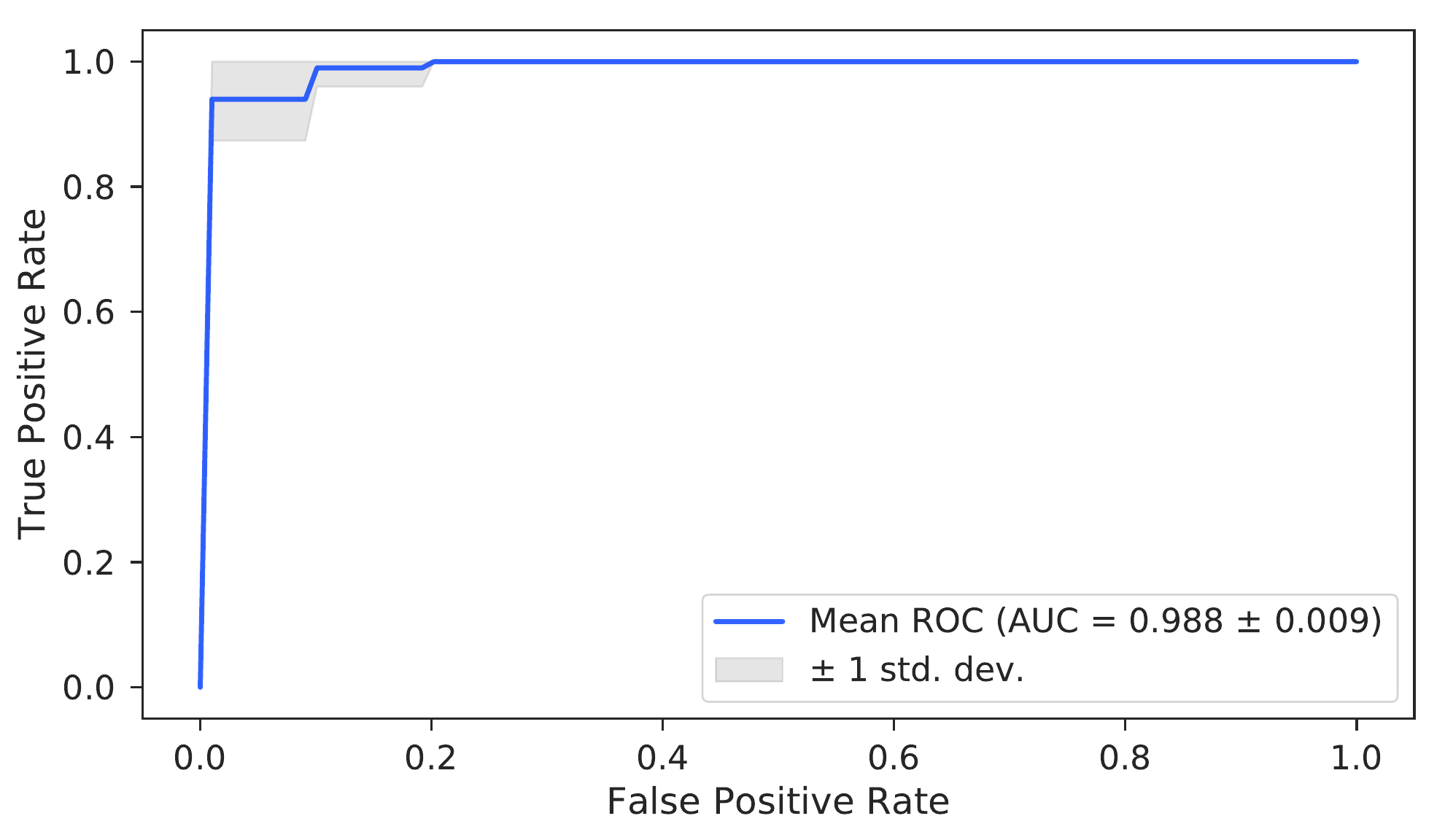}
    \caption{Receiver operating characteristic curve}
    \label{fig:roc_curve}
\end{figure}

We evaluated our model using the 10-fold stratified cross validation applying the Random Forest classifier~\cite{scikit-learn}, using all our selected features. Figure~\ref{fig:roc_curve} shows the Receiver Operating Characteristic (ROC) curve -- the dependency between False Positive (FPR) and True Positive (TPR) Rates. The graph confirms that even with simple statistical dynamic features, it is possible to detect mining activity with high confidence. Indeed, in our experiment we managed to achieve 95\% of accuracy with the Area Under Curve (AUC) score equal to 0.988$\pm$0.009. Our prototype model proves that it is possible to build a very accurate detection tool working at runtime that is able to detect and block mining activities on a device.

We admit that collection of dynamic features is connected with large power consumption overheads. This means that implementation of \tool\ to run on a user device could be impractical. Moreover, the obtained machine learning model is valid in our testbed and may be not transferable to any device. However, the developed approach could be applied during application vetting process~\cite{TrustoreDemo_Zhauniarovich2013}. In this case, app stores could inspect an application and use \tool\ as one of the tests. Indeed, in controlled environment an app could be tested on a particular device, which a priori has a trained model.    


\section{Discussion}
\label{sec:threats}

We have shared our approaches to identify mining code, our set of static miner indicators, and we have presented a technique based on dynamic performance-related features for detecting mining apps on Android. These approaches have certain limitations. 

Static indicators we collected are relevant to our dataset, and new generations of cryptojacking malware apps will likely include different tokens. Moreover, cryptomining libraries evolve or die, and new cryptocurrencies and novel mining algorithms emerge constantly. Yet, our ideas focus on the Android platform specifics that will not change with the cryptomining evolution. It is likely that CPU mining on Android will continue to exist in the \emph{JavaScript} and the \emph{binary} flavors. We have also seen evidence that the low marginal profit from mobile mining drives the attackers to implement the most straightforward and easy solutions. Thus, our suggestions on manual miner analysis and static indicator collection will likely be valid for some time in the future.

Our dynamic detection approach is based on CPU performance profiling. We have demonstrated its practical viability on our dataset. As cryptomining libraries may and will change in the future, our machine learning-based approach will require retraining. However, our approach itself will be relevant, until the underlying core idea of cryptomining, which involves solving computationally hard puzzles quickly (proof-of-work), will change to other consensus mechanisms. 

Our dataset shared with the community contains many apps that rely on the CoinHive library for mining (note that many different mining libraries are also present in our dataset). While CoinHive was shut down in 2019, Monero miners are still a threat\footnote{\url{https://www.bbc.com/news/world-europe-49494927}}. Moreover, at the time of writing, CoinHive is still the dominant browser-based cryptomining library\footnote{\url{https://cryptobriefing.com/is-web-mining-still-a-thing/}}. Therefore, it is important to be able to detect apps with this library, and our results and the apps with this library are still relevant for the community.

\section{Related Work}\label{sec:relwork}
To the best of our knowledge, ours is the first work reporting on Android cryptomining applications. Previously, cryptojacking has been investigated in the context of traditional binary malware \cite{pastrana2019first, huang2014botcoin}, and there have been several papers focusing on browser-based cryptojacking \cite{konoth2018minesweeper,musch2018web,ruth2018digging,eskandari2018first,saad2018end,rauchberger2018other}. 

\subsection{Cryptojacking in Other Contexts.}
\emph{Browser-based cryptojacking.}
The ease of integration of Coinhive-like services into websites has led to the proliferation of drive-by cryptomining attacks. Eskandari et al.~\cite{eskandari2018first} have applied keyword-based search to the website code on the PublicWWW database, and have reported finding more than 30K occurrences of the Coinhive library and some occurrences of its alternatives, such as Crypto-Loot and JSECoin. Konoth et al.~\cite{konoth2018minesweeper} have found 20 active crypto-mining campaigns and 28 crypto-mining services in Alexa's Top 1 Million websites. They have used keyword-based search in web traffic logs, followed by manual analysis. Like in our approach, their keywords included mining services' names and specific strings pertinent to these services (in the miner initialization code and in the Wasm/asm.js, i.e., web assembly, mining payload), Stratum protocol keywords, WebSocket communication. They have also used a high number of WebWorker threads in a web site as a feature pertinent to mining. Similarly, Musch et al.~\cite{musch2018web} report that 0.25\% websites from Alexa Top 1 Million are serving crypto-mining code. They have applied CPU usage profiling and presence of web assembly code and several WebWorkers as indicators. Hong et al.~\cite{hong2018you} have proposed a run-time mining detection tool CMTracker that integrates hash computation-based and stack structure-based profilers. 

Ruth et al.~\cite{ruth2018digging} have seeded their mining website dataset from the NoCoin list~\cite{nocoin} and have proposed a fingerprinting technique for Wasm code. Rauchberger et al.~\cite{rauchberger2018other} have proposed the MiningHunter technique to detect browser-based miners by analysing Web logs.

Saad, Khormali and Mohaisen~\cite{saad2018end} have used public services (Pixalate and Netlab 360) to acquire a list of websites with mining code embedded. Using these sites as ground truth, they have developed dynamic mining script profiles with respect to CPU usage, battery drain and network usage. Machine learning-based approach to browser-based miner detection has also been outlined in Carlin et al.~\cite{carlin2018detecting}, where opcode traces have been used as features, and in Draghicescu et al.~\cite{draghicescu2018crypto}, where the CPU allocation features and the threads and socket connections have been captured. Inlined reference monitoring for Web cryptojackers have been proposed by Wang et al.~\cite{wang2018seismic}.

\emph{Binary-based cryptojacking.}
Malicious Bitcoin cryptominers have been investigated by Huang et al.~\cite{huang2014botcoin} already in 2014. Division into campaigns and profits generated by the recent binary-based cryptominers have been analysed by Pastrana and Suarez-Tangil~\cite{pastrana2019first}. The data collection approach used in~\cite{pastrana2019first} is similar to ours, as the authors crawled public services for malicious samples and then applied static and dynamic analysis heuristics to select only miners. 

In contrast to the aforementioned works related to browser-based and binary-based cryptojacking, ours focuses on the mining applications in the Android ecosystem. Our results show that the Android platform is affected by both Web-based and binary cryptojackers. Yet, we collected and analyzed not only malicious cryptominers, as in \cite{pastrana2019first}, but also bona fide miners that some users might want to explore. We have also reported about the phenomenon of scam miners, that only pretend to be mining cryptocurrencies, while, at best, only serving ads to the users. 

To the best of our knowledge, the SophosLabs report on mining Android apps~\cite{sophos2018} is the only paper analyzing Android mining malware and providing some samples. We have used this report to seed our miner dataset, as it only mentions a few samples hashes (not all samples were obtainable through our main app sources, VirusTotal and Koodous, and available application markets or the app database AndroZoo~\cite{allix2016androzoo}).

\subsubsection{Energy and CPU Consumption Evaluation.}
Recently, Clay et al.~\cite{clay2018power} have evaluated the CPU consumption required for mining on Android devices. As mentioned, Saad, Khormali and Mohaisen~\cite{saad2018end} reported on using CPU usage and battery level on several devices, including an Android phone, to discriminate mining web scripts from non-mining ones (that were emulated with JavaScript disabled in the browser). The authors reported that mining scripts have had significant impact on the CPU and battery (at least 40\% of CPU usage on Android with low throttle and higher rate of battery charge consumption. 

Our dynamic detection approach relies on evaluation of many dynamically profiled features of Android applications, including CPU usage and battery drain caused by computation-intensive mining code. In contrast to \cite{saad2018end,carlin2018detecting}, our solution for dynamic miner detection is based on comparison of mining apps with benign but fully functional ones. 

Several approaches for detecting Android malware based on energy consumption fingerprints have been proposed and evaluated, e.g.,~\cite{merlo2015measuring,hoffmann2013mobile,canfora2016acquiring,caviglione2016seeing,gao2016energy}. Yet, these works focused on detection of malware behaviors other than mining. 


\subsubsection{Android malware detection.}
As our analysis of VirusTotal results and security industry reports show~\cite{fireeyereport2018}, mining functionality can be delivered as a part of malicious payload. There exist a large body of work that focuses on Android malware detection, e.g., \cite{jiang2012dissecting,arp2014drebin,grace2012riskranker,yang2015appcontext,wang2014exploring,xu2016hadm,zhu2016featuresmith,tam2017evolution}, to name just a few. Particularly, the cross-language Dual-Force technique~\cite{tan2015securing} has a big potential for Web-based miner detection. 




	



\section{Conclusions}

Cryptojacking poses a serious threat to mobile devices. At best, illicit cryptocurrency miners deplete the battery of mobile devices fast. However, they may cause more serious damage: from monetary loss to physical harm to the device's owner due to overheating. In order to better comprehend this threat, we collected a dataset of \textbf{728} Android mining apps, and dissected them. To the best of our knowledge, this is the first work that looks into cryptojacking applications on Android. Our analysis confirms the public knowledge in this area is largely insufficient. For example, we found \textbf{173} \emph{illicit} miners from \textbf{76} mining campaigns that have been not previously reported.

In addition, we performed the analysis of the miners from our sample with \emph{VirusTotal}. Our findings are very interesting: \textbf{16} miners from our dataset are not detected by any antivirus engine, and a single miner has been detected by at most \textit{39 out of the total of 74} scanners available at \emph{VirusTotal}, meaning that there is no consistency among the scanner engines. 

With the clean dataset available, we performed a dynamic analysis of the miners and compared the results with benign applications. We identified a set of dynamic metrics that contribute the most to the accurate classification results. Indeed, based on our dataset, we managed to achieve 95\% of accuracy with the AUC score of whoping 0.988$\pm$0.009, according to the 10-fold cross validation. Based on our findings, we proposed a tool called \tool\ that can be used to detect miners at runtime. 

For the future work, we plan to extend our dataset with more illicit cryptocurrency miners that use heavy code obfuscation and to study them. We also plan to extend \tool\ with reliable techniques that account for various static and dynamic evasion methods such as network traffic obfuscation, CPU throttling, and evading user interaction (e.g., mining only when a user does not interact with a device, or at night).

\paragraph{Acknowledgements.}
This research was supported by Luxembourg National Research Fund through grants C15/IS/10404933/COMMA and AFR-PhD-11289380-DroidMod.

\bibliographystyle{acm}
\bibliography{references/short,references/references}

\begin{thebibliography}{10}

\bibitem{allix2016androzoo}
{\sc Allix, K., Bissyand{\'e}, T.~F., Klein, J., and Le~Traon, Y.}
\newblock Androzoo: Collecting millions of android apps for the research
  community.
\newblock In {\em Mining Software Repositories (MSR), 2016 IEEE/ACM 13th
  Working Conference on\/} (2016), IEEE, pp.~468--471.

\bibitem{arp2014drebin}
{\sc Arp, D., Spreitzenbarth, M., Hubner, M., Gascon, H., and Rieck, K.}
\newblock Drebin: Effective and explainable detection of android malware in
  your pocket.
\newblock In {\em Proc. of NDSS\/} (2014), pp.~23--26.

\bibitem{canfora2016acquiring}
{\sc Canfora, G., Medvet, E., Mercaldo, F., and Visaggio, C.~A.}
\newblock Acquiring and analyzing app metrics for effective mobile malware
  detection.
\newblock In {\em Proceedings of the 2016 ACM on International Workshop on
  Security And Privacy Analytics\/} (2016), ACM, pp.~50--57.

\bibitem{carlin2018detecting}
{\sc Carlin, D., O’Kane, P., Sezer, S., and Burgess, J.}
\newblock Detecting cryptomining using dynamic analysis.
\newblock In {\em 2018 16th Annual Conference on Privacy, Security and Trust
  (PST)\/} (2018), IEEE, pp.~1--6.

\bibitem{caviglione2016seeing}
{\sc Caviglione, L., Gaggero, M., Lalande, J.-F., Mazurczyk, W., and
  Urba{\'n}ski, M.}
\newblock Seeing the unseen: revealing mobile malware hidden communications via
  energy consumption and artificial intelligence.
\newblock {\em IEEE Transactions on Information Forensics and Security 11}, 4
  (2016), 799--810.

\bibitem{clay2018power}
{\sc Clay, J., Hargrave, A., and Sridhar, R.}
\newblock A power analysis of cryptocurrency mining: A mobile device
  perspective.
\newblock In {\em 2018 16th Annual Conference on Privacy, Security and Trust
  (PST)\/} (2018), IEEE, pp.~1--5.

\bibitem{CoinHive_Discountinuation}
{\sc {Coinhive}}.
\newblock Discontinuation of coinhive, February 2019.

\bibitem{cyberthreatreport2018}
{\sc {Cyber Threat Alliance}}.
\newblock The illicit cryptocurrency mining threat,
  \url{https://www.cyberthreatalliance.org/wp-content/uploads/2018/09/CTA-Illicit-CryptoMining-Whitepaper.pdf},
  2018.

\bibitem{dashevskyi2020dissecting}
{\sc Dashevskyi, S., Zhauniarovich, Y., Gadyatskaya, O., Pilgun, A., and
  Ouhssain, H.}
\newblock Dissecting {Android} cryptocurrency miners.
\newblock In {\em Proceedings of the Tenth ACM Conference on Data and
  Application Security and Privacy (CODASPY)\/} (2020), ACM.

\bibitem{draghicescu2018crypto}
{\sc Draghicescu, D., Caranica, A., Vulpe, A., and Fratu, O.}
\newblock Crypto-mining application fingerprinting method.
\newblock In {\em 2018 International Conference on Communications (COMM)\/}
  (2018), IEEE, pp.~543--546.

\bibitem{fireeyereport2018}
{\sc Eitzman, R., Goody, K., Wolcott, B., and Kennelly, J.}
\newblock How the rise of cryptocurrencies is shaping the cyber crime
  landscape: The growth of miners,
  \url{https://www.fireeye.com/blog/threat-research/2018/07/cryptocurrencies-cyber-crime-growth-of-miners.html},
  2018.

\bibitem{eskandari2018first}
{\sc Eskandari, S., Leoutsarakos, A., Mursch, T., and Clark, J.}
\newblock A first look at browser-based cryptojacking.
\newblock {\em arXiv\/} (2018).

\bibitem{feldman2014manilyzer}
{\sc Feldman, S., Stadther, D., and Wang, B.}
\newblock Manilyzer: automated android malware detection through manifest
  analysis.
\newblock In {\em Mobile Ad Hoc and Sensor Systems (MASS), 2014 IEEE 11th
  International Conference on\/} (2014), IEEE, pp.~767--772.

\bibitem{gao2016energy}
{\sc Gao, X., Liu, D., Liu, D., and Wang, H.}
\newblock On energy security of smartphones.
\newblock In {\em Proceedings of the Sixth ACM Conference on Data and
  Application Security and Privacy\/} (2016), ACM, pp.~148--150.

\bibitem{grace2012riskranker}
{\sc Grace, M., Zhou, Y., Zhang, Q., Zou, S., and Jiang, X.}
\newblock Riskranker: scalable and accurate zero-day android malware detection.
\newblock In {\em Proceedings of the 10th international conference on Mobile
  systems, applications, and services\/} (2012), ACM, pp.~281--294.

\bibitem{hoffmann2013mobile}
{\sc Hoffmann, J., Neumann, S., and Holz, T.}
\newblock Mobile malware detection based on energy fingerprints—a dead end?
\newblock In {\em International Workshop on Recent Advances in Intrusion
  Detection\/} (2013), Springer, pp.~348--368.

\bibitem{hong2018you}
{\sc Hong, G., Yang, Z., Yang, S., Zhang, L., Nan, Y., Zhang, Z., Yang, M.,
  Zhang, Y., Qian, Z., and Duan, H.}
\newblock How you get shot in the back: A systematical study about
  cryptojacking in the real world.
\newblock In {\em Proc.\ of CCS\/} (2018).

\bibitem{nocoin}
{\sc {hoshsadiq}}.
\newblock Nocoin adblock list
  \url{https://github.com/hoshsadiq/adblock-nocoin-list}, 2019.

\bibitem{huang2014botcoin}
{\sc Huang, D.~Y., Dharmdasani, H., Meiklejohn, S., Dave, V., Grier, C., McCoy,
  D., Savage, S., Weaver, N., Snoeren, A.~C., and Levchenko, K.}
\newblock Botcoin: Monetizing stolen cycles.
\newblock In {\em NDSS\/} (2014), Citeseer.

\bibitem{jiang2012dissecting}
{\sc Jiang, X., and Zhou, Y.}
\newblock Dissecting android malware: Characterization and evolution.
\newblock In {\em Proc.\ of S\&P\/} (2012).

\bibitem{kaspersky-loapi-2017}
{\sc Kaspersky}.
\newblock Loapi -- this trojan is hot!, 2017.

\bibitem{konoth2018minesweeper}
{\sc Konoth, R.~K., Vineti, E., Moonsamy, V., Lindorfer, M., Kruegel, C., Bos,
  H., and Vigna, G.}
\newblock Minesweeper: An in-depth look into drive-by cryptocurrency mining and
  its defense.
\newblock In {\em Proc.\ of CCS\/} (2018).

\bibitem{luo2011attacks}
{\sc Luo, T., Hao, H., Du, W., Wang, Y., and Yin, H.}
\newblock Attacks on webview in the android system.
\newblock In {\em Proc.\ of ACSAC\/} (2011).

\bibitem{ma2010password}
{\sc Ma, W., Campbell, J., Tran, D., and Kleeman, D.}
\newblock Password entropy and password quality.
\newblock In {\em Proc.\ of NSS\/} (2010).

\bibitem{merlo2015measuring}
{\sc Merlo, A., Migliardi, M., and Fontanelli, P.}
\newblock Measuring and estimating power consumption in android to support
  energy-based intrusion detection.
\newblock {\em Journal of Computer Security 23}, 5 (2015), 611--637.

\bibitem{musch2018web}
{\sc Musch, M., Wressnegger, C., Johns, M., and Rieck, K.}
\newblock Web-based cryptojacking in the wild.
\newblock {\em arXiv preprint arXiv:1808.09474\/} (2018).

\bibitem{papadopoulos2018truth}
{\sc Papadopoulos, P., Ilia, P., and Markatos, E.~P.}
\newblock Truth in web mining: Measuring the profitability and cost of
  cryptominers as a web monetization model.
\newblock {\em arXiv preprint arXiv:1806.01994\/} (2018).

\bibitem{pastrana2019first}
{\sc Pastrana, S., and Suarez-Tangil, G.}
\newblock A first look at the crypto-mining malware ecosystem: A decade of
  unrestricted wealth.
\newblock {\em arXiv preprint arXiv:1901.00846\/} (2019).

\bibitem{scikit-learn}
{\sc Pedregosa, F., Varoquaux, G., Gramfort, A., Michel, V., Thirion, B.,
  Grisel, O., Blondel, M., Prettenhofer, P., Weiss, R., Dubourg, V.,
  Vanderplas, J., Passos, A., Cournapeau, D., Brucher, M., Perrot, M., and
  Duchesnay, E.}
\newblock Scikit-learn: Machine learning in {P}ython.
\newblock {\em Journal of Machine Learning Research 12\/} (2011), 2825--2830.

\bibitem{SnapdragonProfiler}
{\sc {Qualcomm Technologies, Inc.}}
\newblock Snapdragon profiler
  \url{https://developer.qualcomm.com/software/snapdragon-profiler}, 2019.

\bibitem{rauchberger2018other}
{\sc Rauchberger, J., Schrittwieser, S., Dam, T., Luh, R., Buhov, D.,
  P{\"o}tzelsberger, G., and Kim, H.}
\newblock The other side of the coin: A framework for detecting and analyzing
  web-based cryptocurrency mining campaigns.
\newblock In {\em Proceedings of the 13th International Conference on
  Availability, Reliability and Security\/} (2018), ACM, p.~18.

\bibitem{ruth2018digging}
{\sc R{\"u}th, J., Zimmermann, T., Wolsing, K., and Hohlfeld, O.}
\newblock Digging into browser-based crypto mining.
\newblock In {\em Proc.\ of IMC\/} (2018).

\bibitem{saad2018end}
{\sc Saad, M., Khormali, A., and Mohaisen, A.}
\newblock End-to-end analysis of in-browser cryptojacking.
\newblock {\em arXiv preprint arXiv:1809.02152\/} (2018).

\bibitem{salem2018repackman}
{\sc Salem, A., Paulus, F.~F., and Pretschner, A.}
\newblock Repackman: a tool for automatic repackaging of android apps.
\newblock In {\em Proceedings of the 1st International Workshop on Advances in
  Mobile App Analysis\/} (2018), ACM, pp.~25--28.

\bibitem{smali}
{\sc smali/backsmali}.
\newblock \url{https://github.com/JesusFreke/smali}, 2018.

\bibitem{sophos2018}
{\sc {Sophos Labs}}.
\newblock Coinminer and other malicious cryptominers targeting android, 2018.

\bibitem{tan2015securing}
{\sc Sufatrio, D. J.~T., Chua, T.-W., and Thing, V.}
\newblock Securing android: a survey, taxonomy, and challenges.
\newblock {\em ACM Computing Surveys (CSUR) 47}, 4 (2015), 58.

\bibitem{tam2017evolution}
{\sc Tam, K., Feizollah, A., Anuar, N.~B., Salleh, R., and Cavallaro, L.}
\newblock The evolution of android malware and android analysis techniques.
\newblock {\em ACM Computing Surveys (CSUR) 49}, 4 (2017), 76.

\bibitem{CoinMiners_ZDNet}
{\sc Tung, L.}
\newblock Android security: Coin miners show up in apps and sites to wear out
  your cpu, October 2017.

\bibitem{wang2018seismic}
{\sc Wang, W., Ferrell, B., Xu, X., Hamlen, K.~W., and Hao, S.}
\newblock Seismic: Secure in-lined script monitors for interrupting
  cryptojacks.
\newblock In {\em European Symposium on Research in Computer Security\/}
  (2018), Springer, pp.~122--142.

\bibitem{wang2014exploring}
{\sc Wang, W., Wang, X., Feng, D., Liu, J., Han, Z., and Zhang, X.}
\newblock Exploring permission-induced risk in android applications for
  malicious application detection.
\newblock {\em IEEE Transactions on Information Forensics and Security 9}, 11
  (2014), 1869--1882.

\bibitem{apktool_webpage}
{\sc Wi\'{s}niewski, R., and Tumbleson, C.}
\newblock {Apktool - A tool for reverse engineering 3rd party, closed, binary
  Android apps.}\url{https://ibotpeaches.github.io/Apktool}, 2017.

\bibitem{xu2016hadm}
{\sc Xu, L., Zhang, D., Jayasena, N., and Cavazos, J.}
\newblock Hadm: Hybrid analysis for detection of malware.
\newblock In {\em Proc.\ of IntelliSys\/} (2016).

\bibitem{yang2015appcontext}
{\sc Yang, W., Xiao, X., Andow, B., Li, S., Xie, T., and Enck, W.}
\newblock Appcontext: Differentiating malicious and benign mobile app behaviors
  using context.
\newblock In {\em Proceedings of the 37th International Conference on Software
  Engineering\/} (2015), pp.~303--313.

\bibitem{StaDynA_Zhauniarovich2015}
{\sc Zhauniarovich, Y., Ahmad, M., Gadyatskaya, O., Crispo, B., and Massacci,
  F.}
\newblock Stadyna: Addressing the problem of dynamic code updates in the
  security analysis of android applications.
\newblock pp.~37--48.

\bibitem{zhauniarovich2016small}
{\sc Zhauniarovich, Y., and Gadyatskaya, O.}
\newblock Small changes, big changes: an updated view on the android permission
  system.
\newblock In {\em Proc.\ of RAID\/} (2016), Springer, pp.~346--367.

\bibitem{TrustoreDemo_Zhauniarovich2013}
{\sc Zhauniarovich, Y., Gadyatskaya, O., and Crispo, B.}
\newblock Demo: Enabling trusted stores for android.
\newblock In {\em Proc.\ of CCS\/} (2013), pp.~1345--1348.

\bibitem{zhu2016featuresmith}
{\sc Zhu, Z., and Dumitras, T.}
\newblock Featuresmith: Automatically engineering features for malware
  detection by mining the security literature.
\newblock In {\em Proceedings of the 2016 ACM SIGSAC Conference on Computer and
  Communications Security\/} (2016), ACM, pp.~767--778.

\end{thebibliography}

\end{document}